\documentstyle[emulateapj,psfig]{article}

\received{}
\accepted{}
\journalid{}{}
\articleid{}{}
\slugcomment{To be published in {\it the Astrophysical Journal}}

\lefthead{KOBAYASHI, TSUJIMOTO, AND NOMOTO}
\righthead{THE HISTORY OF THE COSMIC SUPERNOVAE RATE}

\def\ltsima{$\; \buildrel < \over \sim \;$}
\def\ltsim{\lower.5ex\hbox{\ltsima}}
\def\gtsima{$\; \buildrel > \over \sim \;$}
\def\gtsim{\lower.5ex\hbox{\gtsima}}

\begin{document}

\title{THE HISTORY OF THE COSMIC SUPERNOVA RATE 
DERIVED FROM THE EVOLUTION OF THE HOST GALAXIES}
\author{Chiaki KOBAYASHI$^1$, Takuji TSUJIMOTO$^2$, and Ken'ichi
NOMOTO$^{1,3}$}
\affil{$^1$ Department of Astronomy, School of Science,
University of Tokyo, Bunkyo-ku, Tokyo 113-0033, Japan;
chiaki@astron.s.u-tokyo.ac.jp}
\affil{$^2$ National Astronomical Observatory, 
Mitaka, Tokyo 181-8588, Japan; taku.tsujimoto@nao.ac.jp}
\affil{$^3$ Research Center for the Early Universe, School of Science,
University of Tokyo, Bunkyo-ku, Tokyo 113-0033, Japan;
nomoto@astron.s.u-tokyo.ac.jp}

\begin{abstract}
We make a prediction of the cosmic supernova rate history 
as a composite of the supernova rates in spiral and elliptical galaxies.
We include the metallicity effect on the evolution of 
Type Ia supernova (SN Ia) progenitors, and
construct detailed models for
the evolutions of spiral and elliptical galaxies in clusters and field
to meet the latest observational constraints.
In the cluster environment,
the synthesized cosmic star formation rate (SFR) has 
an excess at $z \gtsim 3$ corresponding
to the early star burst in ellipticals 
and a shallower slope from the present to the
peak at the redshift of $z \sim 1.4$
compared with Madau's plot.  In the field environment, 
we assume that ellipticals form over a wide range of
redshifts as $1 \ltsim z \ltsim 4$.  The synthesized cosmic SFR has a
broad peak around $z \sim 3$, which is in good agreement with the
observed one.  The resultant cosmic SFRs
lead to the following predictions for the cosmic SN Ia rate: 1) The
SN Ia rate in spirals
has a break at $z \sim 2$ due to the low-metallicity inhibition of SNe
Ia, regardless of clusters or field.  2) At high redshifts, the SN Ia
rate has a strong peak around $z \sim 3$ in clusters, whereas in
field much lower rate is expected, reflecting the difference in the
formation epochs of ellipticals.
\end{abstract}


\section{INTRODUCTION}

The search for high-redshift supernovae has been extensively 
conducted mainly to determine cosmological parameters
by the Supernova Cosmology Project (\cite{per99}) 
and the High-z Supernova Search Team (\cite{sch98}). 
High-redshift supernovae can also provide useful information on 
the star formation history in the universe.  
Type Ia supernovae (SNe Ia) 
have been discovered up to $z \sim 1.32$ (\cite{gil99}),
and the SN Ia rate has been estimated 
up to $z \sim 0.5$ (\cite{pai96}; \cite{pai99}).
With the Next Generation Space Telescope, both SNe Ia
and Type II supernovae (SNe II) will be observed through $z \sim 4$.
The SN II rate is directly connected with the star formation rate (SFR),
and the SN Ia rate can also trace the SFR 
combined with a SN Ia progenitor model.
In a theoretical approach, the cosmic SN Ia rate as a function of redshift 
has been constructed using the observed cosmic SFR
(\cite{rui98}; \cite{yun98}, 2000; \cite{sad98}; \cite{mad98}; 
\cite{kob98}, hereafter K98).

The cosmic SFR has been estimated observationally up to $z \sim 5$ from UV
and H$\alpha$ luminosity densities with the help of spectral
population synthesis models (e.g, \cite{mad96}; \cite{con97}).  The
observed cosmic SFR by Madau et al. (1996) shows a peak at 
$z \sim 1.4$ and a sharp decrease to the present.  
However, UV luminosities
which are converted to the SFRs may be affected by the dust extinction
(\cite{pet98a}).  Recent updates of the cosmic SFR show some different
features (\cite{tres98}; \cite{gro98}; \cite{trey98}; \cite{hug98};
\cite{gla99}; \cite{ste99}), one of which suggests that a peak lies
around $z \sim 3$.

Among the several attempts to calculate the cosmic SN Ia rate using the
observed cosmic SFR, K98 predicts that the cosmic
SN Ia rate drops at $z \sim 1-2$, because the SN Ia occurrence 
depends on the metallicity. 
In their SN Ia progenitor model, the accreting 
white dwarf (WD) blows strong winds to reach 
the Chandrasekhar (Ch) mass limit (\cite{hkn96}, 1999b).  
If the iron abundance of the progenitors is as low as
[Fe/H]$\ltsim -1$, then the wind is too weak for SNe Ia to occur.  This
model successfully reproduces the observed chemical evolution of the
solar neighborhood such as the evolution of the oxygen to iron ratio
and the abundance distribution function of disk stars (K98).  

Their finding that the occurrence of SNe Ia depends on the metallicity
of the progenitor systems implies that the SN Ia rate strongly depends
on the history of the star formation and metal-enrichment.  
The universe is composed of different
morphological types of galaxies and therefore the cosmic SFR is a sum
of the SFRs for different types of galaxies.  As each morphological
type has a unique star formation history, we should decompose the
cosmic SFR into the SFR belonging to each type of galaxy and calculate
the SN Ia rate for each type of galaxy.

In this paper, we first construct the detailed evolution models for
different types of galaxies which are compatible with the stringent
observational constraints, and apply them to reproduce the cosmic SFR
for two different environments, i.e., clusters and
field.  Secondly we confirm that the metallicity-dependent SN Ia
progenitor model, which has been already tested for the solar
neighborhood, can explain the present supernova rates for all types of
galaxies.  Finally combining the above SN Ia model with the
self-consistent galaxy models, we calculate the SN Ia rate history for
each type of galaxy and predict the cosmic SN Ia rate as a
function of redshift.

In the next section, we describe our computational method of the
galaxy evolution with our SN Ia progenitor model.  In
section 3, we construct the star formation histories of spirals and
ellipticals, and predict their supernova rate histories in section
4.  In section 5, we make a prediction of the cosmic supernova rates as
a composite of different types of galaxies in clusters and field.
Discussion and conclusions are given in sections 6 and 7, respectively.

\section{MODELS}

\subsection{Type Ia Supernova Model}

The progenitors of the majority of SNe Ia are most likely the Ch mass
WDs (e.g., \cite{nom97a}; \cite{nom00} for recent reviews), although the sub-Ch
mass models might correspond to some peculiar subluminous SNe Ia.  The
early time spectra of the majority of SNe Ia are in excellent
agreement with the synthetic spectra of the Ch mass models, while the
spectra of the sub-Ch mass models are too blue to be compatible with
observations (\cite{hof96}; \cite{nug97}).  For the evolution of
accreting WDs toward the Ch mass, two scenarios have been proposed:
One is a double-degenerate (DD) scenario, i.e., merging of double C+O
WDs with a combined mass surpassing the Ch mass limit (\cite{ibe84};
\cite{web84}), and the other is a single-degenerate (SD) scenario,
i.e., accretion of hydrogen-rich matter via mass transfer from a
binary companion (e.g., \cite{nom94} for a review).  The issue of DD
versus SD is still debated (e.g., \cite{bra95} for a review), but
theoretical modeling has indicated that the merging of WDs does not
make typical SNe Ia (\cite{sai85}, 1998; \cite{seg97}), 
and the lifetime of SNe Ia progenitors
predicted by the DD scenario is too short to be compatible with the chemical
evolution of the solar neighborhood (K98).

Our SD scenario has two progenitor systems: One is a red-giant
(RG) companion with the initial mass of $M_{\rm RG,0} \sim 1 M_\odot$
and an orbital period of tens to hundreds days (Hachisu et al. 1996, 1999ab).
The other is a near main-sequence (MS) companion with an initial mass of
$M_{\rm MS,0} \sim 2-3 M_\odot$ and a period of several tenths of a
day to several days (\cite{li97}; \cite{hknu99}).  
In these progenitor systems, the C+O WD accretes H-rich matter
from the companion star at a sufficiently high rate
($10^{-7}-10^{-6} M_\odot$ yr$^{-1}$) to increase
its mass to the Ch mass through steady hydrogen burning.
With such a rapid accretion, the WD blows optically thick strong winds that
play a key role in stabilizing the mass transfer and avoiding
a common envelope formation.

The optically thick winds are driven by a strong peak of 
OPAL opacity at $\log T ({\rm K}) \sim 5.2$ (e.g., \cite{igl93}). 
Since the opacity peak is due to iron lines, the wind velocity $v_{\rm w}$
depends on the iron abundance [Fe/H] (K98; \cite{hk00}), 
i.e., $v_{\rm w}$ is higher for larger [Fe/H].
The metallicity effect on SNe Ia is clearly demonstrated
by the size of the regions to produce SNe Ia
in the diagram of the initial orbital period versus initial mass 
of the companion star (see Fig.2 of K98).
The SN Ia regions are much smaller for lower metallicity
because the wind becomes weaker.
The wind velocity depends also on the luminosity $L$ of the WD.
The more massive WD has a higher $L$, thus blowing higher velocity winds
(Hachisu et al. 1999b).
In order for the wind velocity to exceed the escape velocity
of the WD near the photosphere,
the WD mass should be larger than a certain critical mass
for a given [Fe/H].
This implies that the initial mass of the WD $M_{\rm WD, 0}$
should already exceed that critical mass in order for
the WD mass to grow to the Ch mass.
This critical mass is larger for smaller [Fe/H], reaching
$1.1 M_\odot$ for [Fe/H] $=-1.1$ (Fig.1 of K98).
Here we should note that the relative number of WDs with
$M_{\rm WD,0} \gtsim 1.1 M_\odot$ is quite small in close binary systems
(\cite{ume99}).
And for $M_{\rm WD,0} \gtsim 1.2 M_\odot$, the accretion leads
to collapse rather than SNe Ia (\cite{nom91}).
Therefore, we assume that no SN Ia occur at [Fe/H] $\le -1.1$
in our chemical evolution model.

The lifetimes of SNe Ia (i.e., the lifetimes of the binary systems from
the formation through the explosion) are determined from the
main-sequence lifetimes of the companion stars
with the initial masses in the range 
$m_{{\rm d},\ell} \le m \le m_{{\rm d},u}$.
In our chemical evolution model, we adopt 
$M_{\rm WD,0}= 1 M_\odot$ and $Z=0.004$
as a representative case.
Then $m_{{\rm d},\ell}=0.9 M_\odot$ and $m_{{\rm d},u}=1.5 M_\odot$
for the RG+WD system and
$m_{{\rm d},\ell}=1.8 M_\odot$ and $m_{{\rm d},u}=2.6 M_\odot$
for the MS+WD system, as seen in Figure 1 of K98.
The distribution function of the companion stars (mass donors) is assumed 
to be $\phi_{\rm d}(m) \propto m^{-x}$ with the slope of $x=0.35$,
and normalized to unity at $m_{{\rm d},\ell} \le m \le m_{{\rm d},u}$.
This slope is determined from the distribution function of 
the initial mass of the companions which is taken from 
the observed mass ratio distribution in binaries (\cite{dug91}).

\begin{figure}
\figurenum{1}
\centerline{\psfig{figure=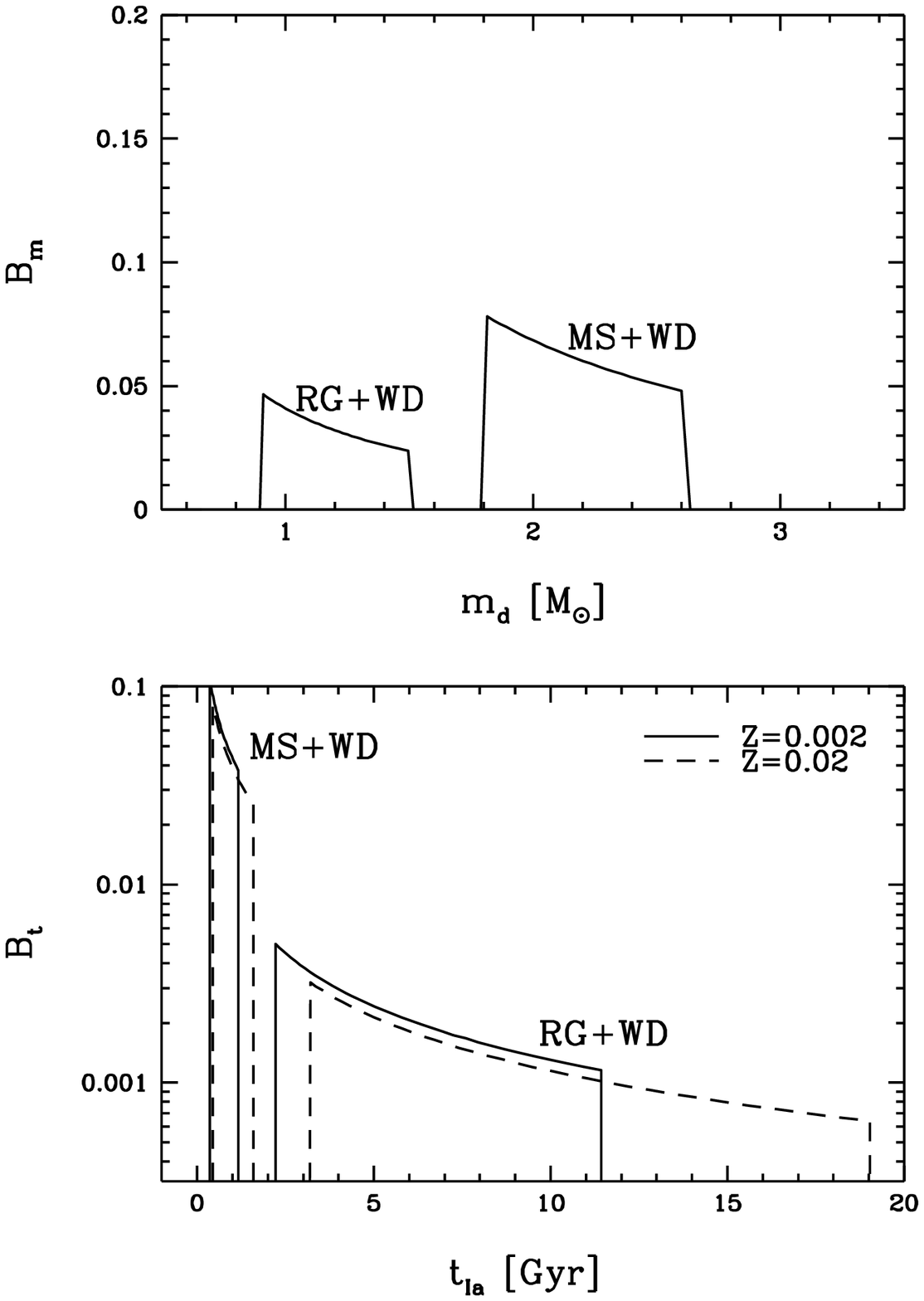,width=18cm}}
\figcaption[fig1.ps]{\label{fig:rate}
The distribution functions of the companion mass (upper panel) 
and the companion lifetime (lower panel). 
In the lower panel, the solid and dashed lines are for 
$Z=0.002$ and $Z=0.02$ (solar), respectively.
MS+WD and RG+WD denote that the companions of the white dwarfs (WD)
is the somewhat evolved near main-sequence (MS) star and the red-giant (RG), 
respectively.}
\end{figure}

The progenitors of SNe Ia are assumed to have main sequence masses 
in the range of $m_{{\rm p},\ell}=3 M_\odot$ and $m_{{\rm p},u}=8 M_\odot$
and form C+O WDs.
The total fraction of primary stars which eventually produce SNe Ia
is denoted by $b$
(see Eq.[\ref{eq:snia}] in section 2.2 for the exact definition of $b$).
The parameter $b$ may depend on the binary fractions.
If the binary fraction depends on the gas density in the 
star forming region, there should be a difference in $b$ between spirals 
and ellipticals (we can expect the higher fraction for ellipticals). 
However the number of low-mass X-ray binaries per unit blue luminosity 
appears to be almost the same between spirals and ellipticals 
(\cite{can87}; \cite{mat94}), which means that the binary fraction 
does not depend on the gas density in the star-forming region.
The parameter $b$ may also depend on the metallicity.
We have not included the metallicity dependence of 
the size of the SN Ia region.
According to the simulations of Hachisu \& Kato (2000),
the area in the orbital period-companion mass plane
does not increase with the metallicity for the MS+WD systems,
but does increase for the RG+WD systems.
Thus, only $b$ for the RG+WD systems $b_{\rm RG}$ 
may be larger for higher metallicity.
However there is no significant difference between
the mean iron abundances of progenitors for the RG+WD systems
in ellipticals and spirals.
Thus, we determine the values of $b$ to reproduce 
the chemical evolution in the solar neighborhood, and
adopt the best-fit values for all types of galaxies.
(We also show the results for other $b$ parameters in Appendix.)
We search the parameter $b$ with the $\chi^2$ test
for the [O/Fe]-[Fe/H] relation and the abundance distribution
in the solar neighborhood,
and get the allowed values of $0.04 \ltsim b_{\rm MS} \ltsim 0.055$ for
the MS+WD systems and $0.01 \ltsim b_{\rm RG} \ltsim 0.04$ for the RG+WD
systems under $2 b_{\rm MS}+b_{\rm RG} \sim 0.12$.
The best-fit values are $b_{\rm MS}=0.05$ and $b_{\rm RG}=0.02$
with the probability of more than $\sim 80\%$; 
the probability is as low as 
$\sim 65\%$ for $b_{\rm MS}=b_{\rm RG}=0.04$\footnote{
In K98, we adopt $b_{\rm MS}=b_{\rm RG}=0.04$ which is
determined from the Kolmogorov-Smirnov (KS) test.} (see Appendix).
The fraction of primary stars which eventually produce SNe Ia is given 
as functions of the companion's mass $m_{\rm d}$ and the companion's lifetime 
(i.e., lifetime of SNe Ia progenitor $t_{\rm Ia}$) as
\begin{equation}
B_m(m_{\rm d}) \equiv b \,
\frac{1}{m_{\rm d}}\,\phi_{\rm d}(m_{\rm d}) \, \left\{
\int_{m_{{\rm d},\ell}}^{m_{{\rm d},u}}\,
\frac{1}{m}\,\phi_{\rm d}(m)~dm \right\}^{-1},
\end{equation}
\begin{equation}
B_t(t_{\rm Ia}) \equiv B_m(m_{\rm d}) \, \frac{d m_{\rm d}}{d t_{\rm Ia}}.
\end{equation}
Figure \ref{fig:rate} shows 
$B_m(m_{\rm d})$ (upper panel) and $B_t(t_{\rm Ia})$ (lower panel)
for the RG+WD and the MS+WD systems.
In the lower panel, the solid and the dashed lines show the cases for 
$Z=0.002$ and $Z=0.02$ (solar), respectively.

In the one-zone uniform model for the chemical evolution of the solar
neighborhood, the heavy elements in the metal-poor stars originate
from the mixture of the SN II ejecta of various progenitor masses.
The abundances averaged over the progenitor masses of SNe II predicts
[O/Fe] $\sim 0.45$ (e.g., \cite{tsu95}; \cite{nom97}).
Later SNe Ia start ejecting mostly Fe, so that [O/Fe] decreases to
$\sim 0$ around [Fe/H] $\sim 0$.  The low-metallicity inhibition of SNe
Ia predicts that the decrease in [O/Fe] starts at [Fe/H] $\sim -1$.
Such an evolution of [O/Fe] well explains the observations (K98).

We should note that some anomalous stars have [O/Fe] $\sim$ 0 at
[Fe/H] $\ltsim -1$.  The presence of such stars, however, is not in conflict
with our SNe Ia models, but can be understood as follows: 
The formation of such anomalous stars (and the diversity of [O/Fe] in
general) indicates that the interstellar materials were not uniformly
mixed but contaminated by only a few SNe II (or even single SN II)
ejecta.  This is because the timescale of mixing was longer than the
time difference between the supernova event and the next generation
star formation.
The iron and oxygen abundances produced by a single SN II vary 
depending on the mass, energy, mass cut, and metallicity of the progenitor.
Relatively smaller mass SNe II ($13-15 M_\odot$) and higher explosion energies
tend to produce [O/Fe] $\sim 0$ (\cite{nom97}; \cite{ume00}).
Those metal poor stars with [O/Fe] $\sim 0$ may be born 
from the interstellar medium polluted by such SNe II.

The metallicity effect on SNe Ia can also be checked with 
the metallicity of the host galaxies of nearby SNe Ia.
There has been no evidence that SNe Ia have occurred
in galaxies with a metallicity of [Fe/H] $\ltsim -1$, 
although host galaxies are detected only for one third of SNe Ia and
the estimated metallicities of host galaxies are uncertain.
Three SNe Ia are observed in low-metallicity dwarf galaxies;
SN1895B and SN1972E in NGC 5253, and SN1937C in IC 4182.
Metallicities of these galaxies are estimated to be
[O/H] $=-0.25$ and $-0.35$, respectively (\cite{koc97}).
If [O/Fe] $\sim 0$ as in the Magellanic Clouds,
[Fe/H]$\sim -0.25$ and $-0.35$ which are not so small.
Even if these galaxies have extremely SN II like abundance as 
[O/Fe] $\sim 0.45$,
[Fe/H] $\sim -0.7$ and $-0.8$ (being higher than $-1$),
respectively.
Since these host galaxies are blue ($B-V=0.44$ for NGC 5253 and
$B-V=0.37$ for IC 4182 according to RC3 catalog), 
the MS+WD systems are dominant progenitors for the present SNe Ia.
The rate of SNe Ia originated from the MS+WD systems is not so sensitive to 
the metallicity as far as [Fe/H] $> -1$ (\cite{hk00}).
Even if [Fe/H] $\sim -0.7$ in such blue galaxies, therefore,
the SN Ia rate is predicted to be similar to those in more metal-rich galaxies.

Futhermore, our SN Ia model can well explain 
the diversity of SN Ia luminosity,
by assuming that the more luminous SNe Ia are produced 
from smaller $M_{\rm WD, 0}$ because of the larger C/O ratio.
Observationally, the most luminous SNe Ia occur only in spirals,
while both spirals and ellipticals are hosts for dimmer SNe Ia (\cite{ume99}).
In our progenitor model, the WD with small $M_{\rm WD, 0}$ ($\sim 0.7 M_\odot$)
can reach the Ch mass only when the companion stars is massive enough
(i.e., $t_{\rm Ia}$ is short enough) to supply
the necessary amount of mass ($\sim 0.7 M_\odot$).
Thus the brightest SNe Ia can occur only near the star forming region.
The variation of the peak brightness of SNe Ia has been found to
depend also on the location in the galaxy, i.e.,
the diversity decreases toward the outer regions of the galaxy (\cite{wan97}).
In our model, this trend is due to the metallicity gradient of the galaxy.
In the outer region, the metallicity is so small that
$M_{\rm WD, 0}$ should be large enough to blow strong winds;
thus only dim SNe Ia appear.

\subsection{Galactic Evolution Model}

A simplified model of the galactic chemical evolution 
is constructed with analytical equations (\cite{tin80}).
The accretion of primordial gas from a reservoir in a halo
is assumed to form a galaxy.
By assuming that the accretion per time $t$
(the infall rate $R_{\rm in}$) is proportional to
the mass of the reservoir, we get 
the exponential form of the infall rate
with a infall timescale $\tau_{\rm i}$ as
\begin{equation}
R_{\rm in}=\frac{1}{\tau_{\rm i}}\exp(-t/\tau_{\rm i}) .
\label{eq:inf}
\end{equation}
Here we define the total baryon mass of the galaxy 
as the mass of the reservoir.

The gas fraction $f_{\rm g}$ is defined as 
the ratio of the gas mass in the bulge and disk
to the total baryon mass of the galaxy (bulge, disk and halo), and
$Z_i$ is the mass fraction of a heavy element $i$ in the gas.
For ellipticals, the gas includes the gas ejected in the galactic wind.
The SFR $\psi$ 
is assumed to be proportional to the gas fraction (\cite{sch59}) as
\begin{equation}
\psi=\frac{1}{\tau_{\rm s}}f_{\rm g} ,
\label{eq:sfr}
\end{equation}
where $\tau_{\rm s}$ is the star formation timescale.
Then the time variations of $f_{\rm g}$ and $Z_i$
are given by the following equations:
\begin{eqnarray}
\frac{df_{\rm g}}{dt}&=&-\psi+E+E_{\rm Ia}+R_{\rm in} ,\\ \label{eq:gas}
\frac{d(Z_if_{\rm g})}{dt}&=&-Z_i\,\psi+E_{z_i}+E_{z_i,{\rm II}}
+E_{z_i,{\rm Ia}}+Z_{i,{\rm in}}\,R_{\rm in} . \label{eq:metal}
\end{eqnarray}
The metallicity $Z$ is defined as the sum of $Z_i$ from C to Zn.
The initial conditions are $f_{\rm g}=0$ and $Z_i=0$.
The metallicity of the infall gas $Z_{i,{\rm in}}$ is assumed to be $0$.

From dying stars, gas is ejected into the interstellar medium
by mass loss and SNe II at a rate of $E$
and by SNe Ia at a rate of $E_{\rm Ia}$.
Heavy elements are ejected at a rate of 
$E_{z_i}$, $E_{z_i,{\rm II}}$, and $E_{z_i,{\rm Ia}}$
by mass loss, SNe II, and SNe Ia, respectively.
These ejection rates are given by the following equations;
\begin{equation}
E=\int_{m_t}^{m_u}\,(1-w_m)\,\psi(t-\tau_m)\,\phi(m)~dm ,
\end{equation}
\begin{equation}
E_{z_i}=\int_{m_t}^{m_u}\,(1-w_m-p_{z_im,{\rm II}})\,Z_i(t-\tau_m)\,
\psi(t-\tau_m)\,\phi(m)~dm ,
\end{equation}
\begin{equation}
E_{z_i,{\rm II}}=\int_{m_t}^{m_u}\,p_{z_im,{\rm II}}\,
\psi(t-\tau_m)\,\phi(m)~dm ,
\end{equation}
\begin{equation}
E_{\rm Ia}=m_{\rm CO}\,{\cal R}_{\rm Ia} ,
\end{equation}
\begin{equation}
E_{z_i,{\rm Ia}}=m_{\rm CO}\,p_{z_im,{\rm Ia}}\,{\cal R}_{\rm Ia} .
\end{equation}
Here $m_{\rm CO}=1.38 M_\odot$ is the white dwarf mass at the explosion.
${\cal R_{\rm Ia}}$ denotes the SN Ia rate,
which is obtained as
\begin{eqnarray}
{\cal R}_{\rm Ia}&=&b~
\int_{\max[m_{{\rm p},\ell},\,m_t]}^{m_{{\rm p},u}}\,
\frac{1}{m}\,\phi(m)~dm~ \nonumber \\ &\times&
\int_{\max[m_{{\rm d},\ell},\,m_t]}^{m_{{\rm d},u}}\,
\frac{1}{m}\,\psi(t-\tau_m)\,\phi_{\rm d}(m)~dm .
\label{eq:snia}
\end{eqnarray}
As noted in section 2.1, our SN Ia scenario has
two types of progenitors (i.e., the MS+WD and the RG+WD systems).
We calculate the SN Ia rate for each binary
with each $b$, $m_{{\rm d},\ell}$, and $m_{{\rm d},u}$, 
and combine them.
The SN II rate ${\cal R_{\rm II}}$ is also obtained as
\begin{equation}
{\cal R}_{\rm II}=\int_{\max[m_{{\rm p},u},\,m_t]}^{m_u}\,
\frac{1}{m}\,\psi(t-\tau_m)\,\phi(m)~dm .
\end{equation}

The lower mass limit for integrals is the turning off mass $m_t$ at $t$ which
is the mass of the star with the main sequence lifetime $\tau_m=t$. 
$\tau_m$ is taken from Kodama \& Arimoto (1997) 
as a function of metallicity $Z$. 
$w_m$ is the remnant mass fraction, 
which is the mass fraction of a neutron star or a white dwarf.  
$p_{z_im,{\rm II}}$ and $p_{z_im,{\rm Ia}}$ are the stellar yields 
which are the mass fractions
of newly produced and ejected heavy element $i$, which are given from 
the supernovae nucleosynthesis model (\cite{tsu95}; \cite{nom97})
with $p_{z_im,{\rm II}}=0$ for $m<10 M_\odot$.
We do not include the dependence of $w_m$,
$p_{z_im,{\rm II}}$ and $p_{z_im,{\rm Ia}}$ on the stellar metallicity.

The initial mass function (IMF) is assumed to have time-invariant mass
spectrum $\phi(m) \propto m^{-x}$ normalized to unity at 
$m_\ell \leq m \leq m_u$.  Theoretical arguments
indicate that the IMF originates from fragmentation of a gas cloud
almost independently of local physics in the gas
(\cite{low76}; \cite{sil77}).  A solar-neighborhood IMF would
therefore be a good approximation, and we adopt the Salpeter slope of
$x=1.35$ (\cite{sal55}) and a mass range from $m_\ell=0.05 M_\odot$ to
$m_u=50 M_\odot$ (\cite{tsu97}).

The time variation of the stellar fraction $f_{\rm s}$ is calculated as
\begin{equation}
\frac{df_{\rm s}}{dt}=\psi-E-E_{\rm Ia} ,
\end{equation}
and the mean stellar metallicity $Z_{i,{\rm s}}$ at the time $t$ 
is obtained from the conservation of heavy elements;
\begin{equation}
Z_i\,f_{\rm g}+Z_{i,{\rm s}}\,f_{\rm s}=
\int_0^t\left\{ E_{z_i,{\rm II}}+E_{z_i,{\rm Ia}}
+Z_{i,{\rm in}}\,R_{\rm in} \right\}\,dt .
\end{equation}

The photometric evolution of galaxies is calculated from the summation
of the simple stellar population, which is defined as a single
generation of coeval and chemically homogeneous stars of various
masses, and taken from Kodama \& Arimoto (1997) as a function of age
$t$ and metallicity $Z$.  The passbands of photometric systems and the
zero points are the same as Kodama \& Arimoto (1997).

For a standard model, 
we adopt $H_0=50$ km s$^{-1}$ Mpc$^{-1}$, $\Omega_0=0.2$, $\lambda_0=0$,
and the galactic age of $15$ Gyr, which corresponds 
to the redshift at the formation epoch of galaxies of $z_{\rm f} \sim 4.5$.

\section{STAR FORMATION HISTORIES OF GALAXIES}

\subsection{Spiral Galaxies}

\begin{figure}
\figurenum{2}
\centerline{\psfig{figure=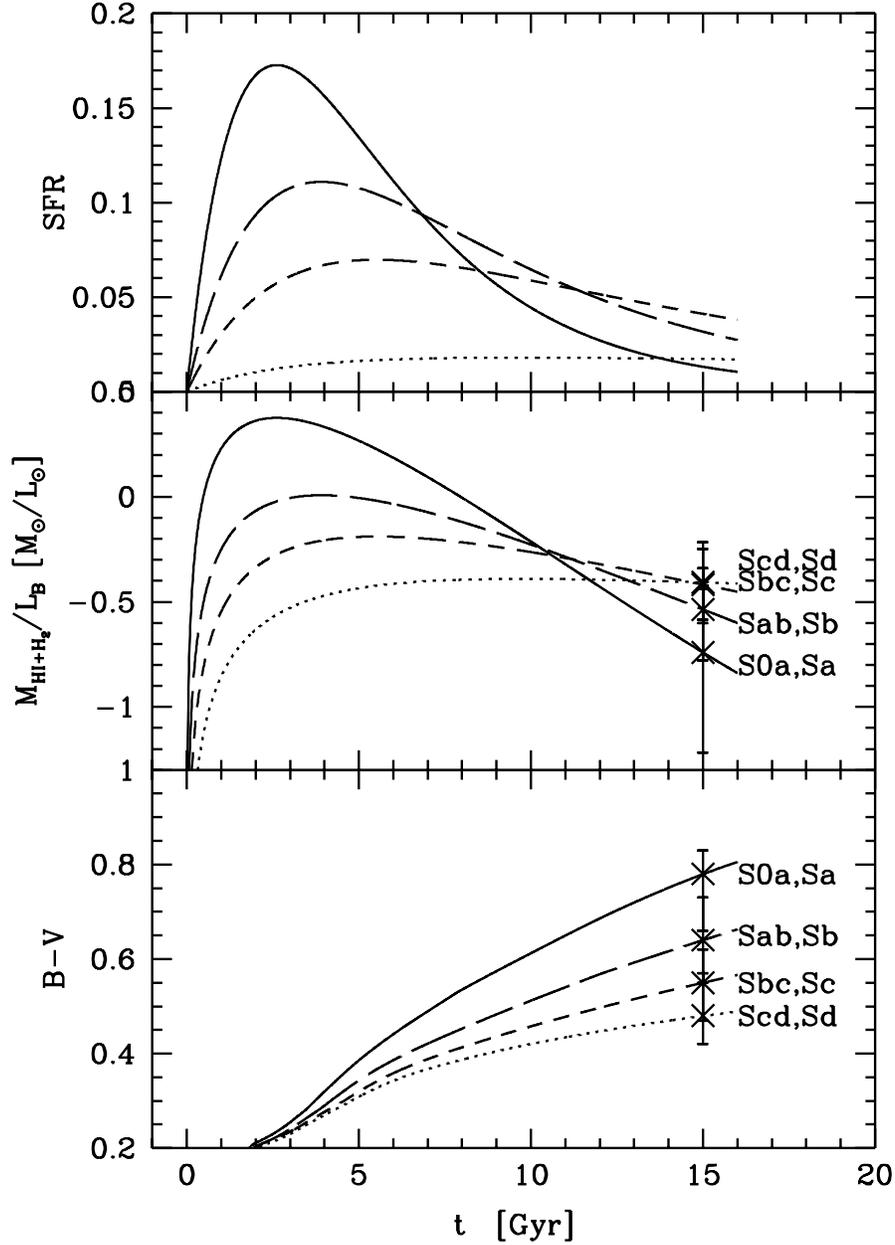,width=18cm}}
\figcaption[fig2.ps]{\label{fig:ssfr}
Star formation rates (SFR: top panel),
gas fractions (middle panel), and $B-V$ colors (bottom panel) 
as a function of time
for four types of spirals : S0a-Sa (solid line), Sab-Sb (long-dashes line), 
Sbc-Sc (short-dashed line), and Scd-Sd (dotted line).
The crosses show the present observed values
of colors and gas fractions.
The present B-V colors are taken from Roberts \& Haynes (1994).
We use the gas (i.e., HI+H$_2$) fractions
which are normalized by the present blue luminosity of the
galaxy to avoid the uncertainty in the fractions of the dark matter.
The HI mass is taken from Roberts \& Haynes (1994), and
the H$_2$ mass is derived from the H$_2$/HI ratios (\cite{cas98}).}
\end{figure}

The star formation history can be inferred from the observed present-day 
colors of galaxies, by using the well-known technique of stellar
population synthesis (\cite{ari92}). 
In this paper,
we determine the timescales $\tau_{\rm i}$ and $\tau_{\rm s}$
in equations (\ref{eq:inf}) and (\ref{eq:sfr})
to reproduce both observed colors and gas fractions
for four types of spirals, i.e., S0a-Sa, Sab-Sb, Sbc-Sc, and Scd-Sd.
These values are summarized in Table \ref{tab:ellin}.
As shown in the top panel of Figure \ref{fig:ssfr},
earlier types of spirals form larger fractions of 
stars at an early epoch, thereby having redder
and smaller gas fractions at present.
The excellent agreements between models and the observations 
are shown in middle and bottom panels of Figure \ref{fig:ssfr}. 
The resultant present metallicities and colors are tabulated 
in Table \ref{tab:ellout}.
Observational data of the present $B-V$ colors for various types of spirals 
are taken from Roberts \& Haynes (1994).
We use the gas (i.e., HI+H$_2$) fractions
which are normalized by the present blue luminosity of the
galaxy to avoid the uncertainty in the fractions of the dark matter.
The HI mass is taken from Roberts \& Haynes (1994), and
the H$_2$ mass is derived from the H$_2$/HI ratios (\cite{cas98}).

\subsection{Elliptical Galaxies}

The star formation history in elliptical galaxies is still controversial,
with single star burst models on one hand and 
continuous star formation  models at the other extreme.
In the former, elliptical galaxies are formed through
dissipative collapse of a protogalactic cloud with a single star burst
at a very early epoch (e.g., \cite{lar74}; \cite{ari87}; \cite{kod97}).  
In the latter, elliptical
galaxies grow through mergers of gaseous galaxies with a continuous
star formation through later epoch (e.g., \cite{kau98}; \cite{bau98}).

The dissipative collapse scenario assumes that the star formation 
in ellipticals has stopped by a loss of gases due to a supernova-driven
galactic wind (\cite{lar74}; \cite{ari87}), so that the bulk of the
stars are old and have formed at $z \gtsim 2$.  
The galactic wind model can well reproduce the passive evolution 
of colors observed in cluster ellipticals (\cite{sta98}; \cite{kod97}).
On the other hand, 
the semi-analytical simulations of hierarchical clustering
(\cite{kau98}; \cite{bau98}) based on the cold dark matter (CDM)
scenario suggest that the formation of ellipticals are protracted
toward lower redshifts especially in the low-density region such
as present-day field.
Such environmental effect is imprinted in
the morphology-density relation (\cite{dre80}; \cite{dre97}).
There are some observational evidences that not
all field ellipticals are as old as cluster ellipticals (\cite{zep97};
\cite{fra98}; \cite{kod99}).  
However the relation of Mg$_2$ indices and velocity dispersions implies
the age difference of only $\sim 1$ Gyr among cluster, group and
field ellipticals (\cite{ber98}).

Taking into account the possibility that there exists the
environmental effect on the galaxy formation, we set two models for
ellipticals in clusters and field, respectively.

\noindent
1) For cluster ellipticals, we adopt the galactic wind model.
The epoch of galactic wind $t_{\rm gw}$
(i.e., the epoch of the end of star formation)
is determined from
the dynamical potential of the galaxy (\cite{lar74}), and we assume
$t_{\rm gw}=1$ Gyr on the average which corresponds to the redshift
of $z \sim 3$. 
The adopted $\tau_{\rm i}$ and $\tau_{\rm s}$ 
in equations (\ref{eq:inf}) and (\ref{eq:sfr})
are summarized in Table \ref{tab:ellin}, and
the resultant present metallicities and colors are 
in Table \ref{tab:ellout}.
This star formation history is constructed to
reproduce the observational constraints such as the present 
mean stellar metallicity of $Z_{\rm s} \sim 0.6 Z_\odot$ 
averaged over the whole galaxy (\cite{koba99}), 
the present $B-V$ color (\cite{rob94}), and the color evolution
observed at $0 \ltsim z \ltsim 1$ as shown in Figure \ref{fig:esfr}a.
The circles, triangles, and crosses show 
the average differences in the observed colors 
of cluster galaxies at each redshifts relative to 
the same rest-frame colors of Coma E+S0 galaxies (\cite{sta98}).
Thus, no color difference represents no evolution relative to Coma.
The observed color differences are as small as our model prediction, 
which means that ellipticals have evolved passively from $z \sim 1$.
In this model, the duration of the star formation is
short enough to be consistent with the observational estimates of 
the magnesium to iron ratio [Mg/Fe] $\sim 0.3$ 
(\cite{wor92}; \cite{koba99}).

\noindent
2) For field ellipticals, we adopt the same star formation
history as in cluster ellipticals, and
assume that the formation epochs span a wide range of redshifts.
The distribution function of the formation epoch $\phi_{z_{\rm f}}$ 
is assumed as
\begin{equation}
\phi_{z_{\rm f}} \propto 
\frac{1}{3}\left(z-\frac{2\sqrt{3}}{\sqrt{3}-1}\right)^3-
\left(\frac{2}{\sqrt{3}-1}\right)^2 
\left(z-\frac{2\sqrt{3}}{\sqrt{3}-1}\right) ,
\end{equation}
which gives
$\phi_{z_{\rm f}}=0$ at $z=0$ and
$\frac{d\phi_{z_{\rm f}}}{dz}=0$ at $z=2$,
and is normalized to unity at $0 \leq z \leq z_{\rm f}$.
As shown in Figure \ref{fig:esfr}b,
this formulation can meet the distribution function of $z_{\rm f}$ 
which is derived from the observational estimates
of $z_{\rm f}$ for 34 field ellipticals in the Hubble Deep Field,
using the broadband spectra (\cite{fra98}).

\section{SUPERNOVA RATES IN GALAXIES}

Present supernova rates observed in the various types of galaxies
(\cite{cap97}; Cappellaro, Evans \& Turatto 1999) 
put the constraints on the SN Ia progenitor
models.  Using the galaxy models in section 3, we show that
our SN Ia model can well reproduce the present supernova rates in both
spirals and ellipticals.

\subsection{Spiral Galaxies}

The observed SN II rate ${\cal R}_{\rm II}$ in late-type spirals is about
twice the rate in early-type spirals.  On the other hand, the observed 
SN Ia rates ${\cal R}_{\rm Ia}$ in both types of spirals are nearly the same.
Thus the present ${\cal R}_{\rm Ia}/{\cal R}_{\rm II}$ ratio in early-type
spirals is larger than that in late-type spirals by a factor of $\sim 2$.  
Such a difference in 
the relative frequency is a result of the difference in the SFR 
(see section 3.1), because the different dependences of ${\cal R}_{\rm II}$ 
and ${\cal R}_{\rm Ia}$ on the SFR are due to the different 
lifetimes of supernova progenitors. Therefore the observed 
${\cal R}_{\rm Ia}/{\cal R}_{\rm II}$ ratio gives a constraint on the SN Ia 
progenitor model.

Figure \ref{fig:ssnr} shows the predicted evolutionary change in 
${\cal R}_{\rm Ia}/{\cal R}_{\rm II}$ ratios
for early and late types of spirals, 
compared with the observations.  
The solid, dashed, and dotted lines show the results for our SN Ia
model, the single delay-time model with $t_{\rm Ia} \sim 1.5$ Gyr 
(Yoshii, Tsujimoto \& Nomoto 1996), 
and the DD model (\cite{tut94}), respectively.  
In the DD model, 
the lifetime of majority of SNe Ia is $\sim 0.1-0.3$ Gyr, 
so that the evolution of ${\cal R}_{\rm Ia}$ is similar to 
${\cal R}_{\rm II}$.  Therefore ${\cal R}_{\rm Ia}/{\cal R}_{\rm II}$ is 
insensitive to the SFR.  This results in the small 
differences in ${\cal R}_{\rm Ia}/{\cal R}_{\rm II}$
among the various type of spirals, which is not consistent
with observations. 
For the similar reason, the single delay-time model with 
$t_{\rm Ia} \sim 1.5$ Gyr is not acceptable.

In our SN Ia model, 
if the iron abundance of progenitors is
[Fe/H]$\gtsim -1$, the occurrence of SNe Ia is determined from the
lifetime of the companions, which is
$t_{\rm Ia} \sim 0.5-1.5$ Gyr for the MS companions 
and $\sim 2-20$ Gyr for the RG companions (see Fig.\ref{fig:rate}).
If SNe Ia occurred only in the MS+WD systems 
with relatively short lifetimes, ${\cal R}_{\rm Ia}/{\cal R}_{\rm II}$ 
would have been insensitive to the SFR. On the contrary, if SNe Ia occurred
only in the RG+WD systems with longer lifetimes, the present difference 
in ${\cal R}_{\rm Ia}/{\cal R}_{\rm II}$ between early
and late type spirals would have been too large, 
reflecting the large difference in SFR at an early epoch.  
Owing to the presence of these two kinds of the 
progenitor systems in our SN Ia progenitor model, the observed 
difference in ${\cal R}_{\rm Ia}/{\cal R}_{\rm II}$ can be reproduced.
We note that the present SN Ia rate depends on the $b$ parameters, 
especially $b_{\rm RG}$ (see Appendix for more details).

\subsection{Elliptical Galaxies}

Figure \ref{fig:esnr} shows the SN Ia rate history in ellipticals.
As noted in section 3.2, we assume that the
bulk of stars in cluster ellipticals are formed at $z \gtsim 3$
and have ages older than $10$ Gyr.
Thus, in the single delay-time model (dashed line) and 
the DD model (dotted line),
the SN Ia lifetimes are too short to reproduce the observed SN Ia rate
at the present epoch.
In ellipticals, the chemical enrichment takes place so early
(see Fig.\ref{fig:global}b)
that the metallicity effect on SN Ia is not effective, 
and therefore the SN Ia rate depends almost only on the lifetime.
Our SN Ia model (solid line) includes the RG+WD systems with 
$t_{\rm Ia}\gtsim 10$ Gyr,
thus well reproducing the present SN Ia rate in ellipticals.

A burst of SNe Ia occur after $\sim 0.5$ Gyr 
from the beginning of the star formation,
because SNe Ia start to occur from the MS+WD systems at
their shortest lifetime.
The second peak of the SN Ia rate appears after $\sim 2$ Gyr, 
because of the onset of SNe Ia from the RG+WD systems.
If we apply the age-redshift relation in Figure \ref{fig:esnr},
these two peaks appear at $z \sim 2.5$ and $\sim 1.5$, respectively.
From $z \sim 0.2$ to $z \sim 0$,
the SN Ia rate gradually decreases by $\sim 40\%$.
Majority of SN Ia progenitors with $Z=0.002$
have already exploded at $z < 0.2$ 
and only more metal-rich SNe Ia occur at $z \sim 0$.
This is because the lifetime $\tau_m$ of the smallest mass 
companions ($0.9 M_\odot$) depends on the metallicity as 
$\tau_m \sim 11$ Gyr for $Z=0.002$ and $\tau_m \sim 19$ Gyr for $Z=0.02$
(see Fig.\ref{fig:rate}).

The decrease in the SN Ia rate from $z \sim 0.2$ to $z \sim 0$ 
depends also on the star formation history in ellipticals and the galactic age.
If we adopt the galactic age of $12$ Gyr, such a decrease in the SN Ia rate 
does not appear.
If ellipticals have undergone the relatively continuous star formation,
as suggested by the hierarchical clustering simulations, 
the SN Ia rate might keep constant to the present.  
The predicted rate for $H_0=50$ km s$^{-1}$ Mpc$^{-1}$ adopted 
in our calculation is 
a little higher than the observations at present.
We can get better fit for $H_0=65$ km s$^{-1}$ Mpc$^{-1}$,
because the absolute value in our model
is independent of the cosmological parameters,
while the observed SN Ia rate is proportional to $H_0^2$. 
The present SN Ia rate also depends on the $b$ parameter, especially
$b_{\rm RG}$ (see Appendix for more details).

\section{COSMIC SUPERNOVA RATE}

Galaxies that are responsible for the cosmic SFR have different
timescales for the heavy-element enrichment, and the occurrence of
supernovae depends on the metallicity therein. Therefore
we calculate the cosmic supernova rate by summing up the supernova 
rates in spirals and ellipticals with the ratio of the relative mass
contribution.
The relative mass contribution of $i$-th type of galaxies is obtained from
the observed relative luminosity proportion 
$c_i=0.215, 0.185, 0.160, 0.275,$ and $0.165$ (\cite{pen76}) 
for ellipticals, S0a-Sa, Sab-Sb, Sbc-Sc, and Scd-Sd, 
respectively,
and the calculated mass to light ratio in B-band $(M/L_B)_i$ 
for each galaxy model,
as given in Table \ref{tab:ellout}.

To compare the observed cosmic SFR,
we convert the predicted cosmic SFR per mass to 
the cosmic SFR per volume by multiplying a constant 
($\rho_c \Omega_{{\rm g}\infty}$) as follows;
\begin{equation}
\psi_{\rm cosmic}\,[M_\odot {\rm yr}^{-1} {\rm Mpc}^{-3}]
=\rho_c \Omega_{{\rm g}\infty} 
\frac{\Sigma_i \, c_i \, (M/L_B)_i \, \psi_i \,[{\rm Gyr}^{-1}]}
{\Sigma_i \, c_i \,(M/L_B)_i}.
\end{equation}
Here $\rho_c$ is the critical density in the cosmology
($\rho_c=\frac{3c^2}{8\pi G}H_0^2$) and 
$\Omega_{{\rm g}\infty}$ is the initial comoving density of the gas
defined by Pei \& Fall (1995).
$\Omega_{{\rm g}\infty}$ is constrained from
the stellar comoving density at present. 
The gas and stellar comoving densities are given by
\begin{equation}
\Omega_{\rm g}=\Omega_{{\rm g}\infty}
\frac{\Sigma_i \, c_i \, (M/L_B)_i \, f_{{\rm g},i}}
{\Sigma_i \, c_i \,(M/L_B)_i},
\end{equation}
\begin{equation}
\Omega_{\rm s}=\Omega_{{\rm g}\infty}
\frac{\Sigma_i \, c_i \, (M/L_B)_i \, f_{{\rm s},i}}
{\Sigma_i \, c_i \,(M/L_B)_i}.
\end{equation}
We adopt $\Omega_{{\rm g}\infty}=3.5 \times 10^{-3} h^{-1}$
to reproduce the present stellar comoving density 
$\log\Omega_{{\rm s}0}=-2.3$ (\cite{fuk98}),
which gives the present gas comoving density of $\log\Omega_{{\rm g}0}=-3.6$.

The cosmic SN II or Ia rate ${\cal R}_{\rm SN}$ 
per luminosity and per volume are given as
\begin{equation}
{\cal R}_{\rm SN,cosmic}\,[(10^{10}L_{B\odot})^{-1}{\rm Century}^{-1}]
=\frac{\Sigma_i \, c_i \, {\cal R}_{{\rm SN},i}\,
[(10^{10}L_{B\odot})^{-1}{\rm Century}^{-1}]}
{\Sigma_i\, c_i},
\end{equation}
\begin{equation}
{\cal R}_{\rm SN,cosmic}\,[{\rm yr}^{-1}{\rm Mpc}^{-3}]
=\rho_c \Omega_{{\rm g}\infty} 
\frac{\Sigma_i \, c_i \, (M/L_B)_i \,{\cal R}_{{\rm SN},i}\,
[M_\odot^{-1}{\rm Gyr}^{-1}]}
{\Sigma_i \, c_i \, (M/L_B)_i}.
\end{equation}
If we adopt larger $\Omega_{{\rm g}\infty}$, 
we get larger $\Omega_{{\rm s}0}$
and thus larger $\psi_{\rm cosmic}$. 
However, the ${\cal R}_{\rm SN,cosmic}$ per luminosity does not depend on
$\Omega_{{\rm g}\infty}$.

\subsection{In Clusters}

First, we make a prediction of the cosmic supernova rates in 
cluster galaxies. We use the galaxy models which are in good agreements 
with the observational constraints as shown in
Figures \ref{fig:ssfr} and \ref{fig:esfr}.
Figure \ref{fig:cluster}a shows the cosmic SFR along redshift
(solid line) as a composite of those in spirals
(long-dashed line) and ellipticals (short-dashed line).
In our galaxy models, ellipticals undergo a star burst at $z \gtsim 3$
and the duration of the star formation is $\sim 1$ Gyr, while spirals undergo
relatively continuous star formation.  Thus, only the SFR in spirals
is responsible for the cosmic SFR at $z \ltsim 2$. 
For comparison, the observed cosmic SFR in field,
so-called Madau's plot
 (\cite{gal95}; \cite{lil96}; \cite{mad96}; \cite{con97}),
is also plotted.
Compared with Madau's plot, the predicted slope
from the present to the peak at $z \sim 1.4$ 
is a little shallower (see also \cite{tot97}).
Also the predicted peak of the SFR in ellipticals at $z \gtsim 3$
is not seen in Madau's plot.

Possible sources of these differences between the
theoretical SFR and observational data might be as follows;
1) Observational uncertainties:
Recent observations have revealed the controversial situation, i.e.,
the slope of the SFR from $z \sim 0$ to $z \sim 1$ is much shallower 
than the Madau's plot according to
Tresse \& Maddox (1998), Treyer et al. (1998), and Gronwall (1998),
while it is steeper according to
Rowan-Robinson et al. (1997) and Glazebrook et al. (1999).
Also the SFR is higher than Madau's plot at $z \gtsim 2$
 (\cite{hug98}; \cite{pet98a})
and continues to be high toward $z \sim 5$ (\cite{ste99}).
2) The star formation at $z \gtsim 3$
may be hidden by the dust extinction (\cite{pet98a}).
3) Ellipticals may have formed at $z \sim 5$ (Totani et al. 1997).
We here propose one more possibility:
4) The theoretical model is constructed for cluster galaxies,
while the observational data are taken from field galaxies.
Then these rates can be different, 
if the star formation histories depends on their environment.
The cosmic SFR histories in field 
is much less clear than in clusters.

Figure \ref{fig:cluster}b shows the cosmic supernova
rates (solid line) as a composite of those in spirals (long-dashed line) and
ellipticals (short-dashed line).  The upper and lower three lines show
the SN II and Ia rates, respectively.
Our SN Ia model is in good agreement with the observed rate at $z \sim 0.5$
(\cite{pai96}; \cite{pai99});
here the number of sample has been increased from 
3 SNe Ia in Pain et al. (1994)
to 38 SNe Ia in Pain (1999),
so that the errorbar has been greatly reduced.
As we will observe supernovae at higher and higher redshift,
the SN Ia rates are predicted to sharply decrease at certain redshifts,
which we denote as $z_{\rm Ia, s}$, $z_{\rm Ia, e}$, and $z_{\rm Ia, tot}$
for in spirals, ellipticals, and all types of galaxies 
(spirals plus ellipticals), respectively.

The SN Ia rate in spirals drops at $z_{\rm Ia, s} \sim 1.9$ 
because of the low-metallicity inhibition of SNe Ia. 
We can test the metallicity effect 
by finding this drop of the SN Ia rate in spirals,
if high-redshift SNe Ia at $z \gtsim 1.5$ and their host galaxies
are observed with the Next Generation Space Telescope.

In ellipticals, the chemical enrichment takes place so early that 
the metallicity is large enough to produce SNe Ia at $z \gtsim 2$. 
Since we assume that the formation epoch of ellipticals is 
$z_{\rm f} \sim 4.5$,
the SN Ia rate in ellipticals decreases at $z_{\rm Ia, e} \sim 2.6$.
This redshift is determined from the shortest lifetime of SNe Ia,
which is $\sim 0.5$ Gyr in our model (Fig.\ref{fig:rate}).
We should note that $z_{\rm Ia, e}$ depends on 
the formation epoch of ellipticals, which is constrained from the
color-magnitude relation in the cluster elliptical galaxies at $z \gtsim 3$
(\cite{kod97}).
The two peaks of SN Ia rates at $z \sim 2.5$ and $z \sim 1.5$ 
come from the MS+WD and the RG+WD systems, respectively (see section 4.2).
Between these peaks, 
the SN Ia rate decreases at $z_{\rm Ia, e} \sim 1.6$.
(Note, however, this drop disappears if we adopt $z_{\rm f}\sim 3$,
because the peak from the MS+WD systems moves to lower redshifts.)
At $z \gtsim 2.5$, the total SN Ia rate decreases as ellipticals at
$z_{\rm Ia, tot} \sim 2.6$.
At $z \ltsim 2$, the total SN Ia rate decreases as spirals at
$z_{\rm Ia, tot} \sim 1.9$, because the contribution of ellipticals 
in the total SN Ia rate is smaller than spirals.
However, the contribution of ellipticals can be larger
depending on $b$ (see Appendix).
Figure \ref{fig:cluster}c is the same as Figure \ref{fig:cluster}b,
but for the supernova rate per volume.

These redshifts, $z_{\rm Ia, s}$, $z_{\rm Ia, e}$ and $z_{\rm Ia, tot}$,
depend on the speed of the chemical enrichment in the host galaxies.
Figure \ref{fig:global}b shows the evolution of the iron abundance
of gases in spirals (long-dashed line) and ellipticals (short-dashed line). 
For ellipticals, the model line is truncated at $z=3.05$, where
ellipticals lose their gases by the galactic wind.
As the chemical enrichment in ellipticals takes place 
in short time,
the iron abundance reaches [Fe/H] $\sim -1$ at $z \sim 4.3$.
Then, the low metallicity effect on SNe Ia does not appear.
In spirals, the iron abundance reaches [Fe/H] $\sim -1$ later at $z \sim 2.3$,
which results in the drop of the SN Ia rate at $z_{\rm Ia, s}\sim 1.9$.

For comparison, we calculate the cosmic supernova rates 
for the cosmic SFR constructed by connecting the 
the observational data (Fig.\ref{fig:global}a).
In Figure \ref{fig:global}a,
the solid line (and the dotted line) is obtained with (and without)
taking into account a correction of the dust extinction (\cite{pet98a}).
The initial gas comoving density is assumed to be 
$\Omega_{{\rm g}\infty}=3.2 \times 10^{-3} h^{-1}$.
The adopted cosmic SFR 
increases from the H$\alpha$ data point at $z=0$ (\cite{gal95}),
makes a peak at $z \sim 1.4$, and decreases towards higher redshifts.
It goes through the HDF UV data points which take into account 
a correction of the dust extinction (\cite{pet98a}).
The corresponding cosmic chemical evolution 
is shown by the dotted line in Figure \ref{fig:global}a.
Because of the lower SFR than in Figure \ref{fig:cluster}a at high redshifts,
[Fe/H] $\sim -1$ is reached at smaller redshift
and the total SN Ia rate decreases at $z_{\rm Ia, tot} \sim 1.6$\footnote{
In the model in Figure 4 of K98,
$z_{\rm Ia, tot}$ is as small as $\sim 1.2$,
because the dust correction was not included 
(like the dotted line in Fig.\ref{fig:global}a)
and $\Omega_{{\rm g}\infty}=2 \times 10^{-3} h^{-1}$ was adopted.}.

\subsection{In the Field}

\begin{figure}
\figurenum{3}
\centerline{\psfig{figure=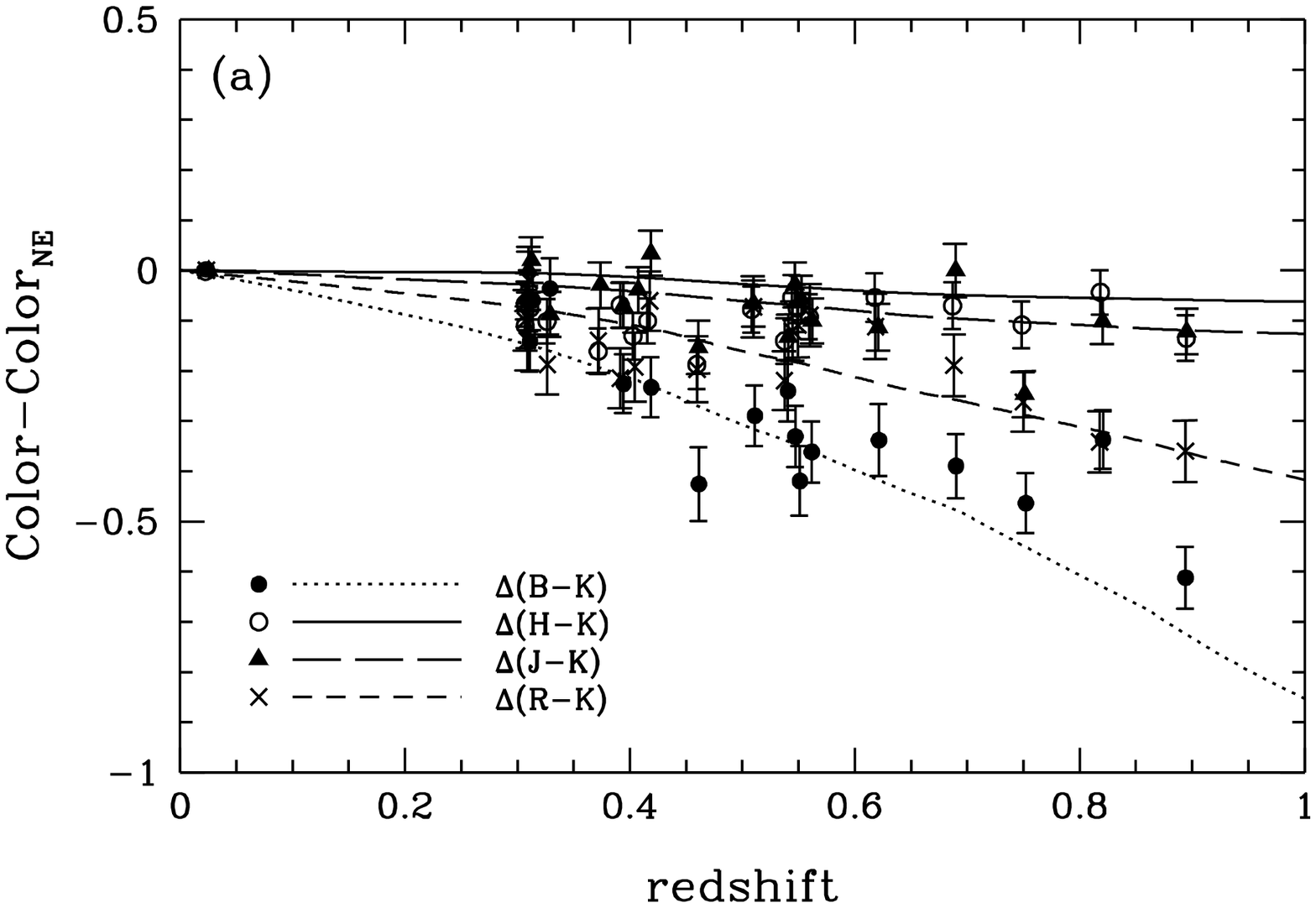,width=13.5cm}}
\centerline{\psfig{figure=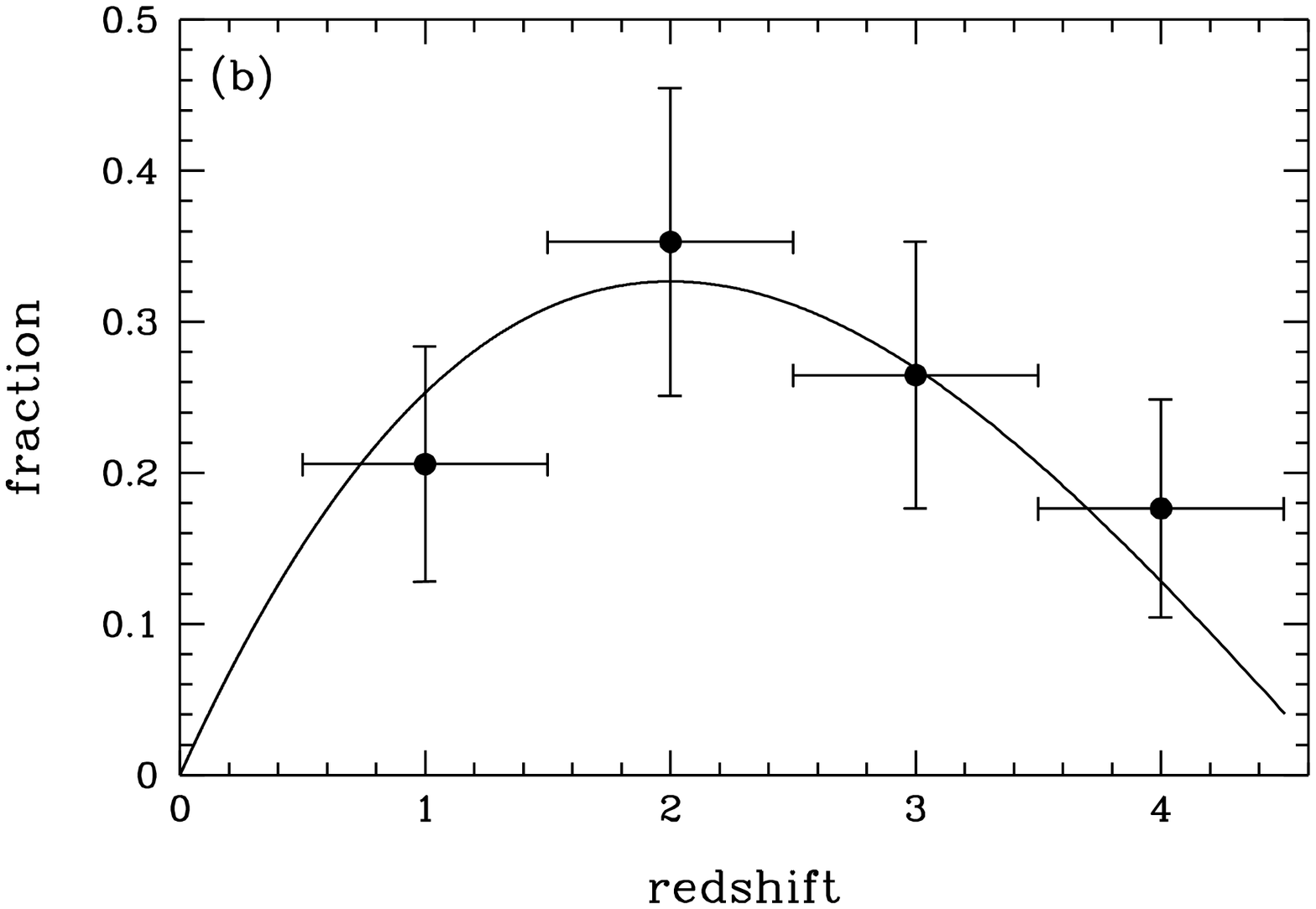,width=13.5cm}}
\figcaption[fig3.ps]{\label{fig:esfr}
(a)~
The passive color evolution predicted by the model of cluster ellipticals 
from the present to $z \sim 1$,
compared with the observational data (\cite{sta98}).
(b)~
The distribution function of the formation epoch
adopted in the model of field ellipticals.
The observational data are estimated
from the spectra of ellipticals in the Hubble Deep Field 
(\cite{fra98}).}
\end{figure}

\begin{figure}
\figurenum{4}
\centerline{\psfig{figure=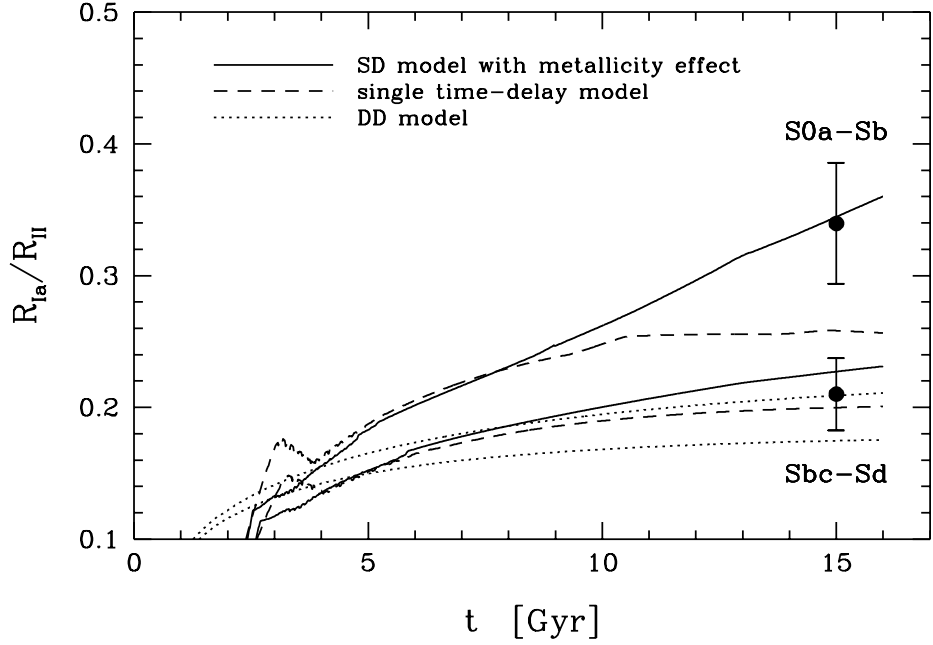,width=13.5cm}}
\caption[fig4.ps]{\label{fig:ssnr}
The ratio of the SN Ia rate to SN II rate ${\cal R}_{Ia}/{\cal R}_{II}$ 
as a function of time for spiral galaxies.
The upper and lower lines show the results for 
the early-type spirals S0a-Sb and for
late-type spirals Sbc-Sd, respectively.
The solid, dashed, and dotted lines are calculated with our SN Ia model, 
the single delay-time model with $t_{\rm Ia} \sim 1.5$ Gyr (\cite{yos96}),
and the DD model (\cite{tut94}), respectively.
The observational data are taken from Cappellaro et al. (1999).}
\end{figure}

\begin{figure}
\figurenum{5}
\centerline{\psfig{figure=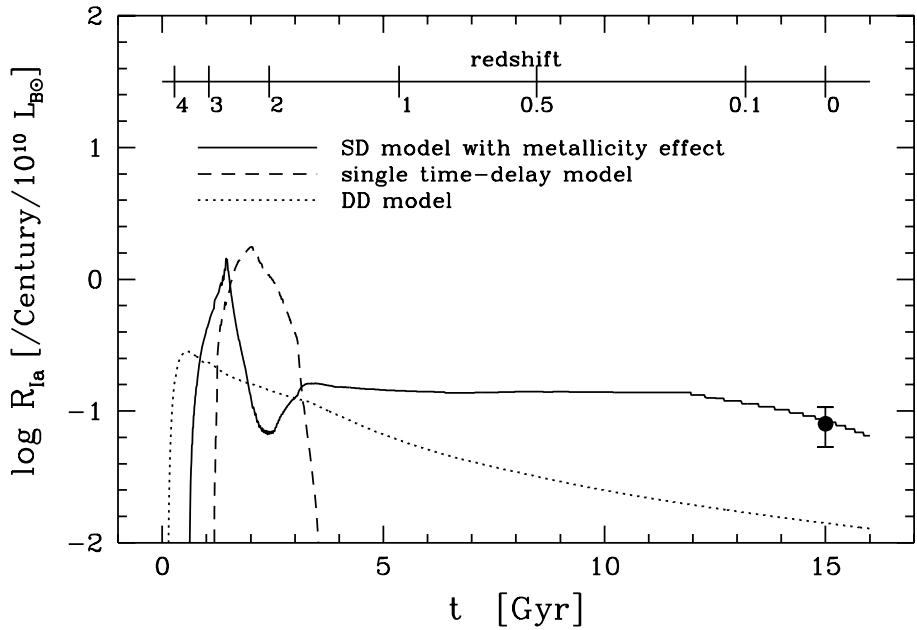,width=13.5cm}}
\caption[fig5.ps]{\label{fig:esnr}
The SN Ia rate as a function of time in elliptical galaxies.
Lines are the same as in Figure \ref{fig:ssnr}.
The observational data is taken from Cappellaro et al. (1999).}
\end{figure}

We also predict the cosmic supernova rates
assuming that the formation of ellipticals in fields 
took place for over the wide range of redshifts
(Fig.\ref{fig:esfr}b).
Figure \ref{fig:field}a shows the cosmic SFR along redshift
(solid line) as a composite of those in spirals (long-dashed line)
and field ellipticals (short-dashed line).
The SFR in spirals is the same as Figure \ref{fig:cluster}a, 
but the star formation in ellipticals continues up to the present.  
Although the timescale of star formation in each elliptical
is as short as in Figure \ref{fig:cluster}a,
the SFR averaged over the field ellipticals gradually increases
toward higher redshifts.
This is because the formation epochs of the field ellipticals
distribute as in Figure \ref{fig:esfr}b.
The synthesized cosmic SFR is in good agreement with
 the observed one, except for the recent H$\alpha$ data at $z \sim 0.9$
(\cite{gla99}).
The peak of the star formation appears at $z
\sim 3$, which is consistent with the recent sub-mm data
(\cite{hug98}).  

Figure \ref{fig:field}b shows the cosmic
supernova rates (solid line) as a composite of those in spirals (long-dashed
line) and ellipticals (short-dashed line).  The upper and lower three
lines show the SN II and Ia rates, respectively.
As in Figure \ref{fig:cluster}b,
the SN Ia rate in spirals drops at $z_{\rm Ia, s} \sim 1.9$. 
The averaged SN Ia rate in ellipticals decreases at $z_{\rm Ia, e} \sim 2.2$ 
as a result of $\sim 0.5$ Gyr delay of the decrease in the SFR at $z \gtsim 3$.
(We should note that $z_{\rm Ia, e}$ depends on 
the distribution function of the formation epochs 
in Figure \ref{fig:esfr}b, which is still uncertain.)
Then, the total SN Ia rate decreases gradually
from $z_{\rm Ia, tot}$ $\sim 2$ to $z \sim 3$.
However, the contribution of ellipticals can be larger
depending on $b$ (see Appendix).
Figure \ref{fig:field}c is the same as Figure \ref{fig:field}b 
but for the supernova rate per volume.

The rate of SNe II in ellipticals evolves 
following the SFR without time delay.
Then, it is possible to observe SNe II in ellipticals around $z \sim 1$.
The difference in the SN II and Ia rates between
cluster and field ellipticals reflects the difference in the
galaxy formation histories in the different environments.

\section{DISCUSSION}

We have made predictions in Figures \ref{fig:cluster}b and \ref{fig:field}b
that the SN Ia rates have breaks at
$z_{\rm Ia, s}$, $z_{\rm Ia, e}$, and $z_{\rm Ia, tot}$
in spirals, ellipticals, and all types of galaxies, respectively.
The exact values of the break redshifts depend on still uncertain factors.
Here we discuss these dependences and uncertainties in order to know 
what we can learn when the Next Generation Space Telescope will obtain
these redshifts.

\subsection{SN Ia Rates in Spiral Galaxies}

For spirals, $z_{\rm Ia, s}$ corresponds to the point where
the metallicity reaches [Fe/H] $\sim -1$.
Therefore $z_{\rm Ia, s}$ is determined by 
the $Z-z$ relation (metallicity-redshift relation), 
which involves the uncertainties in 
the $t-Z$ relation (age-metallicity relation)
and the $t-z$ relation (time-redshift relation).
Note the $t-z$ relation in galaxies depends not only on
the cosmological model but also on 
the formation epoch of the galaxies $z_{\rm f}$.

Among the above factors, the uncertainty in modeling the evolution
of spiral galaxies is relatively small.
The standard evolution model for spiral galaxies has been
obtained by constructing the detailed model for the Galaxy which
has a number of observational constraints. 
Based on the parameters on the IMF and the SN Ia rate 
determined for the Galaxy, 
we have developed the models for other types of spiral galaxies.
Futhermore, the parameters on the SFR have been constrained from the
gas fractions and colors of present-day spirals 
as in Figure \ref{fig:ssfr}.

As for the $t-z$ relation, 
a set of the cosmological parameters ($H_0$, $\Omega_0$, $\lambda_0$)
and $z_{\rm f}$ are constrained
from the Galactic age which settles down to $\sim 12-15$ Gyr from
the recent {\it HIPPARCOS} data (\cite{gra97}; \cite{har97}). 
(Here we assume the same $z_{\rm f}$
for all types of spirals.) In our calculations, we adopt
$(H_0,\Omega_0,\lambda_0)=(50,0.2,0)$ that gives the cosmological age
of $\sim 16.5$ Gyr and thus allows the galactic age of $t_{\rm gal}=15$ Gyr;
these parameters lead to $z_{\rm Ia, s} \sim 1.9$.
If we adopt  $(H_0,\Omega_0,\lambda_0)=(50,1.0,0)$ and $(65,0.2,0)$,
which  give shorter cosmological ages
and allows the galactic age of $t_{\rm age}=12$ Gyr, 
we obtain smaller values of $z_{\rm Ia, s} \sim$ 
$1.3$ and $1.8$, respectively. 
On the other hand, the $\Lambda$-dominant model, e.g.,
$(H_0,\Omega_0,\lambda_0)=(65,0.2,0.8)$
with the galactic age of $t_{\rm age}=15$ Gyr, predicts
$z_{\rm Ia, s} \sim 2.1$. The uncertainty in the $t-z$ relation
therefore results in $1.5 \ltsim z_{\rm Ia, s} \ltsim 2$. 
The uncertainty from the $t-Z$ relation is smaller.

\subsection{SN Ia Rates in Elliptical Galaxies}

For cluster ellipticals, 
the uncertainty in the $t-Z$ relation is much smaller
than spirals, because [Fe/H] 
reaches $\sim -1$ in several $10^8$ years by the star burst.
The parameters for SFR is constrained from the
color evolution of the cluster ellipticals, although
other star formation histories may be allowed for the field ellipticals.

Like $z_{\rm Ia, s}$, the $t-z$ relation affects $z_{\rm Ia, e}$.
Especially a large uncertainty lies in $z_{\rm f}$.
In our calculations, we assume $z_{\rm f} \sim 4.5$ and it results in
the decrease of the SN Ia rate twice at 
$z_{\rm Ia, e} \sim 1.6$ and $\sim 2.6$,
which respectively correspond to 
the lifetimes of the MS+WD and the RG+WD systems
after $z_{\rm f}$ (Fig.\ref{fig:cluster}b).
However, the observational constraint from the color
evolution on $z_{\rm f}$ only requires $z_{\rm f}\gtsim 3$
for cluster ellipticals.
With other cosmological models and $z_{\rm f}$,
the uncertainty in the $t-z$ relation
results in $z_{\rm Ia, e} \gtsim 2$ and
$1 \ltsim z_{\rm Ia, e} \ltsim 2$ for
the MS+WD and the RG+WD systems, respectively.

For field ellipticals, there are large uncertainties in both $t-z$ and
$t-Z$ relations owing to little information from observations,
including the possibility that there is little difference between
cluster and field ellipticals. 
To discuss quantitatively the
uncertainties in $z_{\rm Ia, e}$ in field, 
further observational information would be needed.

\subsection{Total SN Ia Rates}

If the many of the host galaxy types are not identified,
the total rate of the spirals and elliptical components
would depend on the parameters in more complicated manner.

In Figure \ref{fig:cluster}b, the first break of the total SN Ia rate 
appears clearly at $z_{\rm Ia, tot} \sim 1.9$.
This is the case if $z_{\rm f} \sim 4.5$ and
the contribution of ellipticals ($z_{\rm Ia, e} \sim 1.6$) 
is smaller than spirals ($z_{\rm Ia, s} \sim 1.9$).
If $z_{\rm f} \sim 3$, for example, 
the break at $z_{\rm Ia, tot} \sim 1.9$ disappears.
The second break at $z_{\rm Ia, tot} \sim 2.6$ corresponds to
the second $z_{\rm Ia, e}$.

Furthermore the ambiguity in the relative mass fractions $c_i$ of
spirals and ellipticals makes a prediction of the first $z_{\rm Ia, tot}$ 
more complicated, while the second one remains unchanged. 
If the relative contribution from ellipticals is larger, the first
$z_{\rm Ia, tot}$ appears at the first $z_{\rm Ia, e}$
 (e.g., $z \sim 1.6$ for $z_{\rm f} \sim 4.5$). 
Besides in our calculations the relative mass
fractions $c_i$ of galaxies are assumed to be constant against the
redshift. If there is the number evolution of galaxies as suggested by
the hierarchical clustering scenario, the star formation history and
thus the supernova rate in ellipticals are similar to those for field
ellipticals.

\section{CONCLUSIONS}

We have made predictions of the cosmic supernova rate history as a
composite of the supernova rates in different types of galaxies.  We
adopt the SN Ia progenitor scenario including the metallicity effect
which successfully reproduces the chemical evolution of the solar
neighborhood (K98). 
We construct the evolution models for
spiral and elliptical galaxies to meet the latest observational
constraints such as the present gas fractions and colors for spirals,
and the mean stellar metallicity and the color evolution from the
present to $z \sim 1$ for ellipticals.

Owing to the presence of the {\it two} kinds of 
the SN Ia progenitor systems (MS+WD and RG+WD) with 
shorter ($0.5-1.5$ Gyr) and longer lifetimes ($2-20$ Gyr), respectively,
we can explain the difference in the
relative ratio of the SN Ia to SN II rate ${\cal R}_{\rm Ia}/{\cal
R}_{\rm II}$ between early and late types of spirals.  Owing to
the presence of the RG+WD systems with over $10$ Gyr lifetime, SNe Ia
can be seen even at present in ellipticals where the star formation
has already ceased more than 10 Gyr before.

We then construct the cosmic SFR as a composite of the SFRs for
different types of galaxies, and predict the cosmic supernova rates
for two different environments:

\begin{enumerate}
\item 
In the cluster environment, the synthesized cosmic SFR has an excess
at $z \gtsim 3$ due to the early star burst in ellipticals and a
shallower slope from the present to the peak at $z \sim 1.4$, compared
with Madau's plot.  The predicted cosmic supernova rate suggests that
in ellipticals SNe Ia can be observed even at high redshifts because
the chemical enrichment takes place so early that the metallicity is
large enough to produce SNe Ia at $z \gtsim 2.5$. In spirals the SN Ia
rate drops at $z \sim 2$ because of the low-metallicity inhibition of
SNe Ia.
\item 
In the field environment, ellipticals are assumed to form at such a
wide range of redshifts as $1 \ltsim z \ltsim 4$.  The synthesized
cosmic SFR has a broad peak around $z \sim 3$, which is in good
agreement with the recent sub-mm observation by Hughes et al.~(1998). The SN
Ia rate is expected to be significantly low beyond $z \gtsim 2$ because
the SN Ia rate drops at $z \sim 2$ in spirals and gradually decreases
from $z \sim2 $ in ellipticals.
\end{enumerate}

\acknowledgments

This work has been supported in part by the grant-in-Aid for
Scientific Research (08640336) and COE research (07CE2002) of the
Ministry of Education, Science, Culture, and Sports in Japan.  
C.K. thanks to the Japan Society for Promotion 
of Science for a financial support. 
K.N. thanks to the participants in the workshop on
``Type Ia Supernova'' (Aspen Center for Physics, 13-24 June 1999)
for informative discussion.
We would like to thank I. Hachisu and M. Kato for 
providing us with their new results, 
B. Schmidt for fruitful discussions, 
and T. Kodama for providing us with the database of
simple stellar population spectra.

\appendix

\section{$b$ parameter dependence}

\subsection{Present supernova rates}

The SN Ia rate prediction involves some uncertainties 
associated with the SN Ia progenitor model,
which are mostly reflected in the uncertainty in $b$ parameters 
given in section 2.1.
$b_{\rm MS}$ and $b_{\rm RG}$ are the parameters for
the total fraction of the WDs (with the initial mass of $3-8 M_\odot$)
which eventually produce SNe Ia 
in the MS+WD and the RG+WD systems, respectively
(see Eq.[\ref{eq:snia}] in section 2.2 for the exact definition of $b$).
In the present study, the parameter set of
$[b_{\rm MS}=0.02, b_{\rm RG}=0.05]$ is chosen to best reproduce
the chemical evolution in the solar neighborhood.
Other sets of $b$ parameters can also be approximately 
consistent with the chemical evolutions, but
provide different predictions of
the present supernova rates as follows.

\noindent
a) {\it supernova rates in spiral galaxies} ----
In Figure \ref{fig:ssnr},
we use the relative ratio of the present SN Ia to SN II rates
${\cal R}_{\rm Ia}/{\cal R}_{\rm II}$
to avoid the $H_0$ uncertainty included in the observed supernova rates.
The adopted set of $[b_{\rm MS}=0.02, b_{\rm RG}=0.05]$ 
is seen to give also the best fit to ${\cal R}_{\rm Ia}/{\cal R}_{\rm II}$.
If we adopt larger $b_{\rm RG}$,
${\cal R}_{\rm Ia}/{\cal R}_{\rm II}$ is larger.
For $[b_{\rm MS}=0.04, b_{\rm RG}=0.04]$, 
${\cal R}_{\rm Ia}/{\cal R}_{\rm II} \sim$ $0.47$ and $0.26$
for early and late type spirals, respectively,
which are larger than the observed values (\cite{cap99}).
For $[b_{\rm MS}=0.01, b_{\rm RG}=0.055]$, 
${\cal R}_{\rm Ia}/{\cal R}_{\rm II} \sim$ $0.28$ and $0.21$
for early and late type spirals, respectively,
which are smaller.
However, these sets of $b$ parameters may be allowed
in view of the possible
systematic errors and the inhomogeneity of the samples.
We need further updates of the present supernova rates to test our model.

\noindent
b) {\it SN Ia rate in elliptical galaxies} ----
In ellipticals, the star formation has stopped more than $10$ Gyr before and
only SNe Ia from the RG+WD system occur at present (Fig.\ref{fig:esnr}).
Thus $b_{\rm RG}$ can be constrained from 
the present SN Ia rate in ellipticals.
For example, if we adopt the sets of $[b_{\rm MS}=0.04, b_{\rm RG}=0.04]$
and $[b_{\rm MS}=0.055, b_{\rm RG}=0.01]$,
the predicted SN Ia rates are by a factor of $\sim 2$ higher and lower
than the observed values (\cite{cap99}), respectively.
However, such differences from the observed rates
are comparable to the uncertainty stemming from $H_0$,
so that these sets cannot be precluded
until more accurate $H_0$ is obtained.

\subsection{Supernova rates at high and intermediate redshifts}

The supernova rates at high and intermediate redshifts can also 
be used to test our SN Ia model.
In particular, 
whether the $b$ parameters are different among the different types of galaxy
could be seen from the evolution of the supernova rates with redshift.
Observationally, we have found no significant differences in
the binary fraction and the iron abundance between ellipticals and spirals.
Theoretically,
as noted in section 2.1,
$b$ may depend on the binary fraction and the iron abundance; especially,
the metallicity dependence of $b_{\rm RG}$ is significant.
Here we show the results for the cases that $b_{\rm RG}$ 
is larger only for ellipticals.

\begin{figure}
\figurenum{6}
\centerline{\psfig{figure=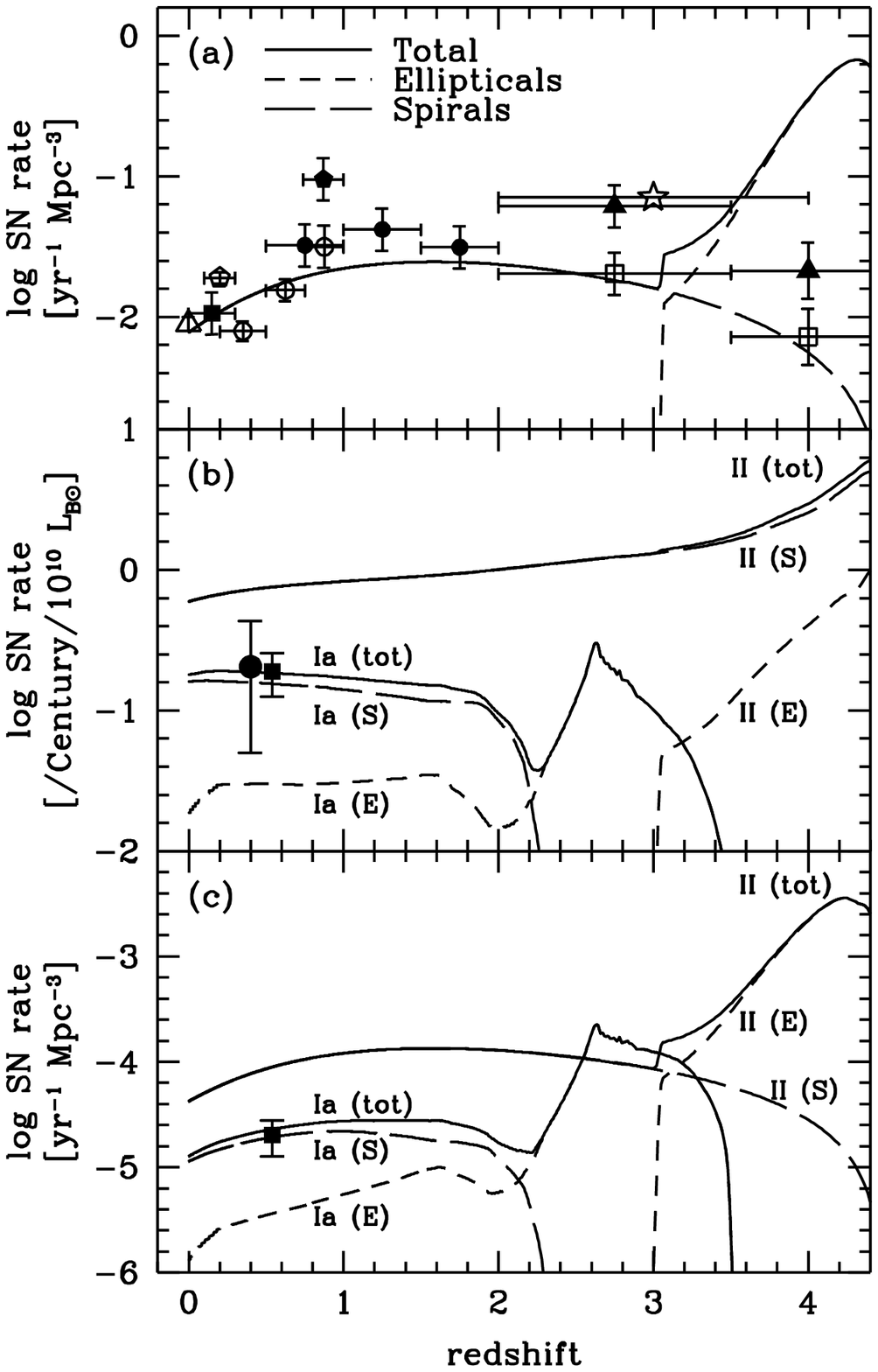,width=19cm}}
\figcaption[fig6.ps]{\label{fig:cluster}
(a)~
The cosmic SFR along redshift (solid line) 
as a composite of those 
in spirals (long-dashed line) and ellipticals (short-dashed line).
The symbols are the observational data for the cosmic SFR
(\cite{gal95}, open triangle; 
\cite{lil96}, open circles; \cite{mad96}, open squares; 
\cite{con97}, filled circles; \cite{tres98}, open pentagon; 
\cite{trey98}, filled square; \cite{gla99}, filled pentagon; 
\cite{hug98}, star; \cite{pet98a}, filled triangle).
(b)~
The cosmic supernova rate (solid line) 
as a composite of those 
in spirals (long-dashed line) and ellipticals (short-dashed line).
The observational data for the cosmic SN Ia rate are taken from 
Pain et al. (1996; circle) and Pain (1999; square).
(c)~
The cosmic supernova rate per volume (solid line)
as a composite of those 
in spirals (long-dashed line) and ellipticals (short-dashed line).
The observational data for the cosmic SN Ia rate is taken from Pain (1999).}
\end{figure}

\begin{figure}
\figurenum{7}
\centerline{\psfig{figure=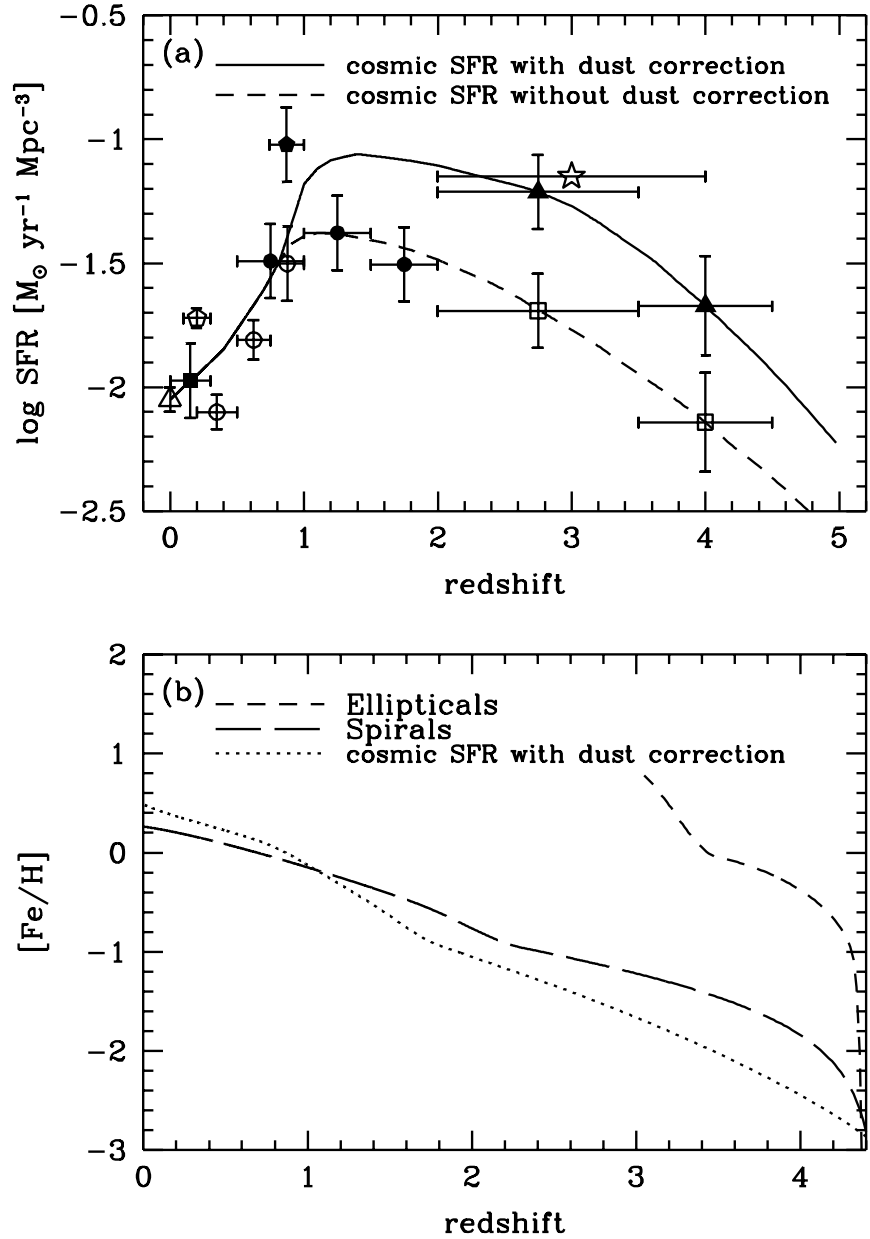,width=18cm}}
\figcaption[fig7.ps]{\label{fig:global}
(a)~
The cosmic SFR along redshift 
which is a connected line of the observed data.
The solid line takes into account the correction of the dust extinction
(\cite{pet98a}) and the dotted line does not.
(b)~
The iron abundance
in the gas of spirals (long-dashed line) and ellipticals (short-dashed line).
For the dotted line,
the observed cosmic SFR with the dust correction is adopted.}
\end{figure}

\begin{figure}
\figurenum{8}
\centerline{\psfig{figure=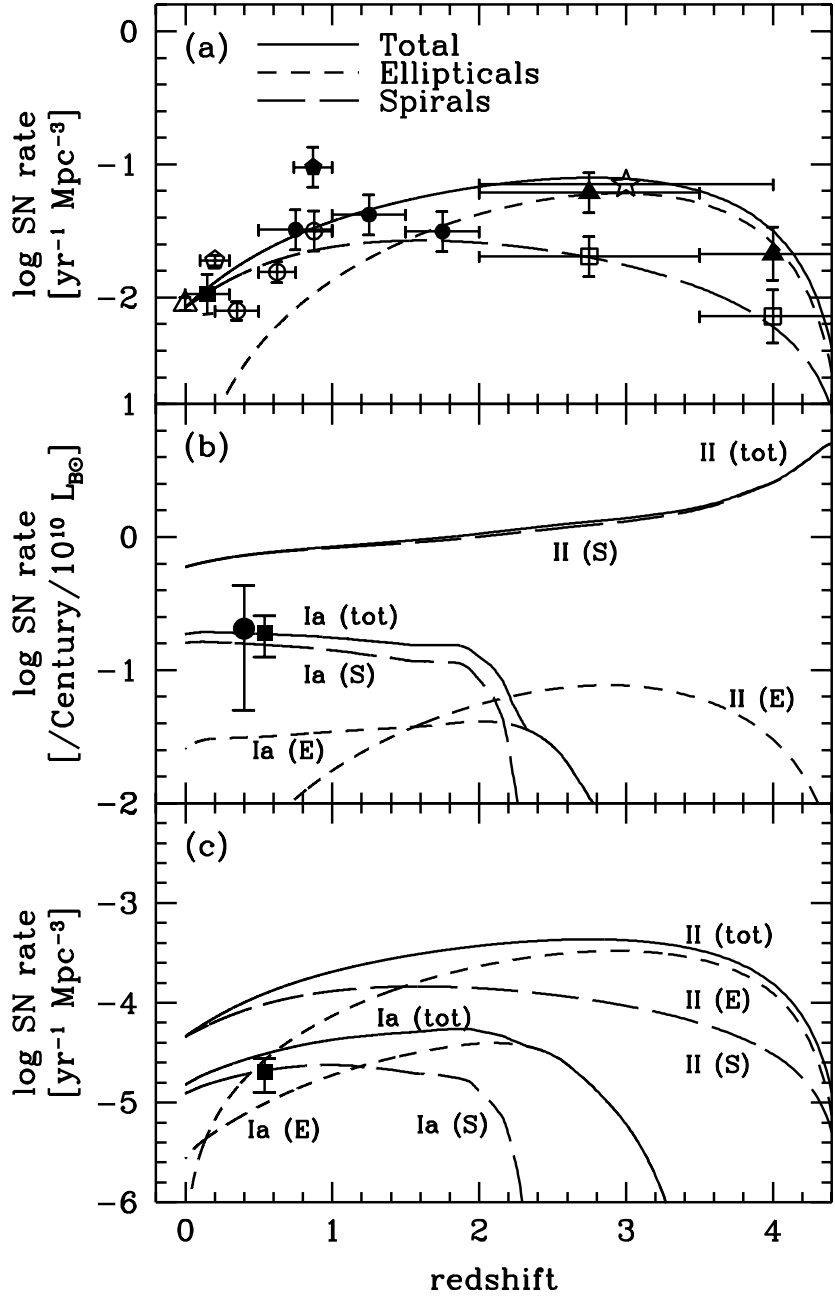,width=19cm}}
\figcaption[fig8.ps]{\label{fig:field}
The same as Figure \ref{fig:cluster}, but for ellipticals 
whose formation epochs span over $1 \ltsim z \ltsim 4$;
this might correspond to field ellipticals.}
\end{figure}

Figure 9 shows the cosmic supernova rates 
for the set of $[b_{\rm MS}=0.05, b_{\rm RG}=0.02]$
in spirals and $[b_{\rm MS}=0.05, b_{\rm RG}=0.10]$ in ellipticals.
In this case the contribution of ellipticals to the cosmic SN Ia rate
is as large as that of spirals.
The redshifts where the SN Ia rates drop are the same
as in Figures \ref{fig:cluster}b and \ref{fig:field}b.
This is because 
$z_{\rm Ia, e}$ of ellipticals is determined by
the lifetime of SNe Ia (i.e, the time-redshift relation)
and not the iron abundance of the galaxy (i.e., the age-metallicity relation,
which is effective only for spirals),
while the difference in $b$ affects only the latter relation.
Whether the contribution of ellipticals is actually as large as spirals
will be seen soon in the observations for the intermediate redshifts
(B. Schmidt, private communication).

In the above case,
the present SN Ia rate in ellipticals is $\sim 5$ times larger 
than the observed rate (\cite{cap99}).
However, the SN Ia rate
could significantly decrease to the present,
if the lowest mass of the RG companions to produce SNe Ia
is not $0.9 M_\odot$ (with the lifetime of $\sim 19$ Gyr for $Z=0.02$)
but $1.0 M_\odot$ (with the lifetime of $\sim 12$ Gyr for $Z=0.02$),
because the star formation in ellipticals
has stopped $14$ Gyr before in our model.
Whether the longest lifetime of SNe Ia is actually
younger than the age of ellipticals
can be seen from the intermediate redshift observations.


\newpage

\begin{figure}
\figurenum{9}
\centerline{\psfig{figure=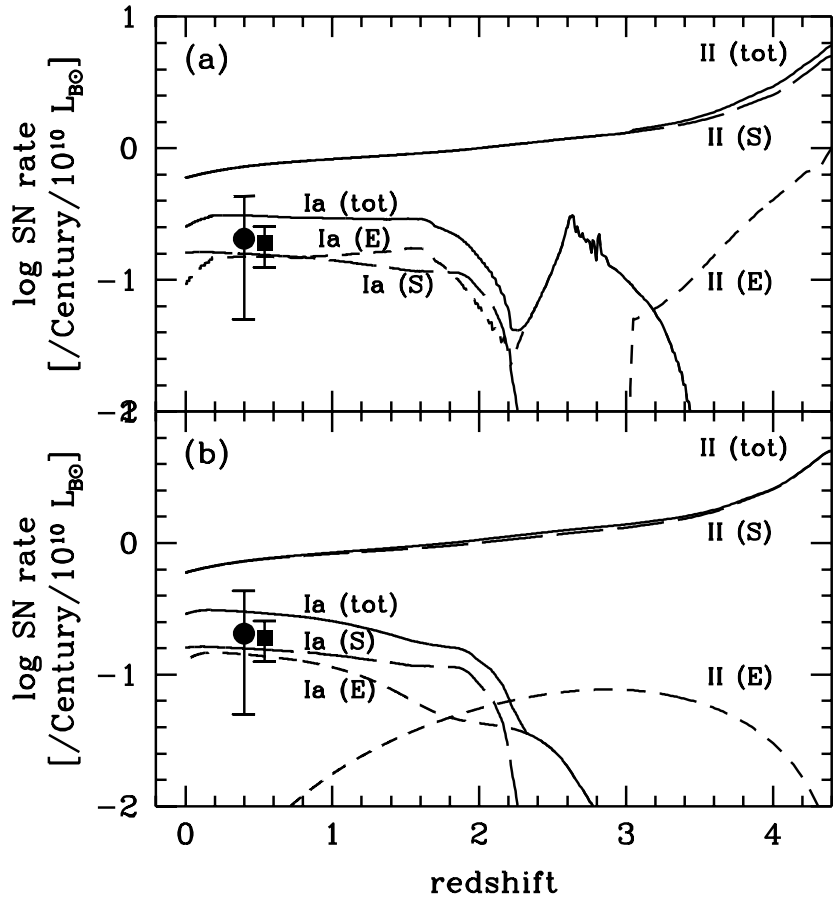,width=19cm}}
\figcaption[fig9.ps]{
The cosmic supernova rate along redshift
in clusters (a) and field (b)
with $[b_{\rm MS}=0.05, b_{\rm RG}=0.02]$ for spirals and 
$[b_{\rm MS}=0.05, b_{\rm RG}=0.10]$ for ellipticals.
Symbols are the same as in Figure \ref{fig:cluster}b.}
\end{figure}

\clearpage

\begin{deluxetable}{lcc}
\tablenum{1}
\footnotesize
\tablewidth{0pt}
\tablecaption{\label{tab:ellin}
The input parameters in equations (\ref{eq:inf}) and (\ref{eq:sfr}).
}
\tablehead{
\colhead{} &
\colhead{$\tau_{\rm s}$ [Gyr]} &
\colhead{$\tau_{\rm i}$ [Gyr]} \\
\colhead{(1)} & \colhead{(2)} & \colhead{(3)}
}
\startdata
E       & $0.1$  & $0.1$  \nl
S0a, Sa & $2.32$  & $2.35$  \nl
Sab, Sb & $2.19$  & $5.61$  \nl
Sbc, Sc & $2.39$  & $10.92$   \nl 
Scd, Sd & $2.19$  & $60.5$   \nl
\enddata
\tablenotetext{}{Col.(1).---Galaxy type.}
\tablenotetext{}{Col.(2)(3).---Timescales of the star formation and inflow 
in Gyr.}
\end{deluxetable}

\begin{deluxetable}{lrrrrrrrr}
\tablenum{2}
\footnotesize
\tablewidth{0pt}
\tablecaption{\label{tab:ellout}
The calculated quantities for spirals and ellipticals at $15$ Gyr.}
\tablehead{
\colhead{} & 
\colhead{$f_{\rm g}$}  &  \colhead{$f_{\rm s}$}  &
\colhead{[M/H]$_{\rm g}$} & \colhead{[M/H]$_{\rm s}$} & 
\colhead{[Fe/H]$_{\rm g}$} & \colhead{[Fe/H]$_{\rm s}$} & 
\colhead{[Mg/Fe]$_{\rm g}$} & \colhead{[Mg/Fe]$_{\rm s}$} \\ 
\colhead{} &  
\colhead{${\cal R}_{\rm II}$} & \colhead{${\cal R}_{\rm Ia}$} & 
\colhead{${\cal R}_{\rm II}$} & \colhead{${\cal R}_{\rm Ia}$} & 
\colhead{$B-V$} & \colhead{$U-B$} &
\colhead{$M/L_B$} & \colhead{$M/L_V$} \\
\colhead{(1)} &
\colhead{(2)} & \colhead{(3)} & \colhead{(4)} & \colhead{(5)} & 
\colhead{(6)} & \colhead{(7)} & \colhead{(8)} & \colhead{(9)} \\
\colhead{} &   
\colhead{(10)} & \colhead{(11)} & \colhead{(12)} & \colhead{(13)} &
\colhead{(14)} & \colhead{(15)} & \colhead{(16)} & \colhead{(17)} \\
}
\startdata
E       &
$ 0.095$ & $ 0.905$ & $ 0.176$ & $-0.212$ & $ 0.831$ & $-0.411$ & $-0.777$ & $ 0.339$ \nl 
& $ 0.000$ & $ 0.007$ & $ 0.000$ & $ 0.084$ & $ 0.924$ & $ 0.339$ & $11.323$ & $ 8.799$ \nl 
S0a, Sa &
$ 0.031$ & $ 0.968$ & $ 0.323$ & $-0.160$ & $ 0.607$ & $-0.026$ & $-0.215$ & $-0.032$ \nl 
& $ 0.072$ & $ 0.034$ & $ 0.425$ & $ 0.198$ & $ 0.776$ & $ 0.106$ & $ 5.880$ & $ 5.238$ \nl 
Sab, Sb &
$ 0.070$ & $ 0.862$ & $ 0.082$ & $-0.184$ & $ 0.315$ & $-0.050$ & $-0.151$ & $-0.032$ \nl 
& $ 0.174$ & $ 0.051$ & $ 0.712$ & $ 0.209$ & $ 0.631$ & $-0.056$ & $ 4.105$ & $ 4.178$ \nl 
Sbc, Sc &
$ 0.098$ & $ 0.648$ & $-0.033$ & $-0.218$ & $ 0.164$ & $-0.092$ & $-0.109$ & $-0.023$ \nl 
& $ 0.225$ & $ 0.053$ & $ 0.853$ & $ 0.202$ & $ 0.541$ & $-0.137$ & $ 3.796$ & $ 4.199$ \nl 
Scd, Sd &
$ 0.038$ & $ 0.182$ & $-0.124$ & $-0.235$ & $ 0.051$ & $-0.112$ & $-0.081$ & $-0.019$ \nl 
& $ 0.095$ & $ 0.019$ & $ 0.953$ & $ 0.194$ & $ 0.472$ & $-0.189$ & $10.085$ & $11.880$ \nl 
\enddata
\tablenotetext{}{Col.(1).---
Galaxy types.}
\tablenotetext{}{Col.(2)(3).---
Gas and stellar fractions.
For ellipticals, $f_{\rm g}$ includes the gas ejected in the galactic wind.}
\tablenotetext{}{Col.(4)(6)(8).---
Gas metallicities [M/H]$_{\rm g}\equiv\log Z/Z_{\odot}$,
iron abundances of gas [Fe/H]$_{\rm g}$, and
magnesium to iron ratios of gas [Mg/Fe]$_{\rm g}$.}
\tablenotetext{}{Col.(5)(7)(9).---
Mean stellar metallicities [M/H]$_{\rm s}\equiv\log Z_{\rm s}/Z_{\odot}$,
mean stellar iron abundances [Fe/H]$_{\rm s}$, and
mean stellar magnesium to iron ratios [Mg/Fe]$_{\rm s}$.}
\tablenotetext{}{Col.(10)(11).---
SN II and Ia rates per mass in the unit of [/Gyr/$10^{3}M_\odot$].}
\tablenotetext{}{Col.(12)(13).---
SN II and Ia rates per luminosity in the unit of [/Century/$10^{10}L_{B\odot}$].}
\tablenotetext{}{Col.(14)(15).---
$B-V$ and $U-B$ colors.}
\tablenotetext{}{Col.(16)(17).---
Mass-to light ratios in B- and V-band. }
\end{deluxetable}

\end{document}